\documentclass[prd,twocolumn,aps]{revtex4-1}

\usepackage{graphicx}
\usepackage{bm}
\usepackage{amsmath}
\usepackage{natbib}
\usepackage{dcolumn}% Align table columns on decimal point
\def\bea {\begin{eqnarray}}
\def\eea {\end{eqnarray}}
\def\be {\begin{equation}}
\def\ee {\end{equation}}
\def\ben{\begin{enumerate}}
\def\een{\end{enumerate}}
\def\bi{\begin{itemize}}
\def\ei{\end{itemize}}

\def\hyphen{{\mbox{-}}}

\def\2p1h{2p\hyphen 1h }
\def\3p2h{3p\hyphen 2h }

\begin{document}

% FRONT MATTER %
%%%%%%%%%%%%%%%%

% TITLE %
%%%%%%%%%

\title{Limits on tensor type weak currents from nuclear and neutron $\beta$~decays}

% AUTHORS %
%%%%%%%%%%%

\author{F. Wauters}
\affiliation{Department of Physics, University of Washington, Seattle, Washington 98195, USA}

\author{A. Garc\'{i}a}
%\email{agarcia3@u.washington.edu}
\affiliation{Department of Physics, University of Washington, Seattle, Washington 98195, USA}

\author{R. Hong}
\affiliation{Department of Physics, University of Washington, Seattle, Washington 98195, USA}

\date{\today}% It is always \today, today, but you may specify any date with \date.

% ABSTRACT %
%%%%%%%%%%%%

\begin{abstract}
The limits on time-reversal invariant tensor-type weak currents from nuclear and neutron $\beta$~decays are evaluated including most recent experimental data. We find that $ -0.14 \times 10^{-2} < (C_T + C^\prime_T)/C_A < 1.4 \times 10^{-2} $ and $ -0.16 < (C_T - C^\prime_T)/C_A < 0.16$ (90\%~C.L.), while for the case $C_T = C^\prime_T$ the limits are $0.0 \times 10^{-2} < C_T/C_A < 0.4 \times 10^{-2}$. These limits are shown to be more stringent than those from recent measurements of the radiative pion decay. In addition, the sensitivity of future $10^{-3}$-level correlation measurements is investigated.
\end{abstract}

%%%%%%%%%%%%%%%%%%%%%%%%%%%%%
%
%PACS numbers
%
%%%%%%%%%%%%%%%%

% PACS, the Physics and Astronomy Classification Scheme.

%%%%%%%%%%%%%%%%%%%
% BODY OF ARTICLE %
%%%%%%%%%%%%%%%%%%%

\maketitle
%\section{\label{sec:level1}Introduction}

\section{Introduction}
The standard model of particle physics prescribes that electro-weak interactions are mediated by vector and axial-vector currents.  However, several theories dealing with deficiencies  of the Standard Model predict chirality-violating scalar and tensor currents. An example are lepto-quarks which are predicted by many unification models and could provide a natural explanation for the remarkable cancelation between the quark and lepton contributions to triangle anomalies~\cite{Buchmuller1987442}. Other ideas, such as left-right symmetric models, aimed at a natural explanation for parity violation~\cite{Herczeg2001}, and supersymmetric models~\cite{profumo2007}\cite{Belyaev1992} also scalar and tensor currents. The $V-A$ structure of the weak currents was established in the late 1950's by measurements of nuclear $\beta$~decay correlations. More recently precision measurements of nuclear and neutron beta decays~\cite{Severijns2006,Behr2009,Abele2008,Dubbers:2011}, pion decays~\cite{Pocanic201026c}, and searches at collider experiments have been pursued in search for non-VA components of the weak currents. 

There are reasons to believe that chirality violating interactions, if they exist, are far smaller than current limits. Some of these arguments involve considerations of the fact that they would not be renormalizable if they existed at the fundamental level. Instead, they could come as the result of exchanges of, for example, supersymmetric particles but these scenarios yield tiny contributions \cite{Belyaev1992}. Other limits come from considering the necessary contributions of chirality-violating interactions to neutrino masses\cite{Ito2005}. Both of these are model dependent. In this paper we only consider direct limits coming from kinematic observables.

Recently, the authors of refs.~\cite{Bhattacharya:12,Cirigliano2013,CiriglianoGardnerHolstein:13} have presented a unified Effective Field Theory framework to compare low- and high energy probes in search for exotic weak currents. Under the assumption that the new physics emerges at a higher energy scale than the production threshold of the LHC, low- and high-energy probes can be compared in a rather model-independent way. It is shown there that presently low-energy experiments are competitive in constraining tensor or scalar currents that couple to left-handed neutrinos. For tensor currents they indicate that the limits coming from radiative pion decays~\cite{bychkov2009} are more stringent than those coming from nuclear beta decays.

In this article we present limits on tensor-type weak currents from nuclear and neutron $\beta$~decays, taking into account recent experimental data and show that the combined limits from nuclear beta decays are actually more stringent than those from pion decays. Note that limits from from neutron $\beta$ decays alone have recently been presented in ref.~\cite{Pattie2013}.

\section{Formalism}

Following Jackson, Treiman, and Wyld~\cite{Jackson1957} we parameterize the nuclear $\beta$ decay hamiltonian in terms of coupling constants $C_i$ and $C_i^\prime$ for each of the possible $V,\; A,\;S,$~and~$\;T$ currents
\footnote{
We neglect pseudo-scalar currents because their role is minimal due the slow nature of the nucleon motion within the nucleus
}:
\begin{eqnarray}
\label{eq:VAST}
H_{int}=\sum_{i=V,A}
(\bar\psi_p O^i \psi_n)\;
\left(
(C_i+C_i^\prime)\; \bar\psi_e^L O_i \psi_\nu^L+ \right. \nonumber \\
\left.
(C_i- C_i^\prime)\; \bar\psi_e^R  O_i \psi_\nu^R
\right)\nonumber \\
+
\sum_{i=S,T}
(\bar\psi_p O^i \psi_n)\;
\left(
(C_i+C_i^\prime)\; \bar\psi_e^R O_i \psi_\nu^L+ \right. \nonumber \\
\left.
(C_i- C_i^\prime)\; \bar\psi_e^L O_i \psi_\nu^R
\right)
\end{eqnarray}
where the operators are:
\begin{eqnarray}
\label{eq:operators}
O_S &=& 1 \nonumber\\
O_P &=& \gamma^5\nonumber\\
O_V &=& \gamma_\mu \nonumber\\
O_A &=& i \gamma_\mu \gamma^5 \nonumber\\
O_T &=& \sigma_{\mu \nu} /\sqrt{2}=-i \left(\gamma_\mu \gamma_\nu - \gamma_\nu \gamma_\mu \right)/(2\sqrt{2}).
\end{eqnarray}
We separated the sum in the hamiltonian of Eq.~\ref{eq:VAST} into the vector and axial-vector parts, allowed in the standard model and conserving chirality, and the scalar and tensor parts, which violate chirality. We restrict our present analysis to only left-handed components for the standard model currents, i.e. assume $C_V=C_V^\prime$ and $C_A=C_A^\prime$ and concentrate on searching for non-zero values for $C_S,\;C_S^\prime$ and $C_T,\;C_T^\prime$.

The distribution $W$ for electrons~(positrons) and (anti)~neutrinos derived from Eq.~\ref{eq:VAST} is given by
\begin{eqnarray}
\label{eqn:jtw}
W \propto \Bigl[ 1 &+ \frac{m_e}{E_e}b_{Fierz}  + A_{\beta} \frac{\bm p_e}{E_e} \cdot  \frac{\bm J}{J} \nonumber \\
&+ a_{\beta\nu} \frac{\bm p_e}{E_e } \cdot \frac{\bm p_\nu}{E_\nu} + \; \cdot\cdot\cdot \; \Bigr]\;,
\end{eqnarray}
with $m_e$ the electron rest mass, ${\bm p_e}\; ({\bm p_{\nu}}$) and $E_e\; (E_{\nu})$ the electron (neutrino) momentum and total energy, and ${\bm J}$ the parent nuclear spin. The dependencies of the correlation coefficients $b_{Fierz}$, $A_{\beta}$, $a_{\beta \nu}$, etc. on the coupling constants in Eq.~\ref{eq:VAST} are listed in~\cite{Jackson1957a}.

%Precision measurements of pion and kaon decay observables can also be used to search for tensor currents. Bolotov et al.\cite{Bolotov:1990} took data from $\pi^- \rightarrow e^-\; {\hat \nu}\;\gamma$ over a wide range of the kinematic variables and found a discrepancy with the expected distribution which they interpreted as possible evidence for Tensor currents following Ref.~\cite{Poblaguev:1990}. Additionally the PIBETA collaboration found distortions in the high-$E_\gamma$ low-$E_{e^+}$ kinematic region of $\pi^+ \rightarrow e^+\; \nu\;\gamma$ \cite{Frlez:2004} which could also be explained away assuming Tensor contributions to the weak decay. However, more recently the PIBETA collaboration published more precise data with a wider kinematic coverage and showed good agreement with the Standard Model expectations\cite{bychkov2009}. The upper limit from this last measurement was used by Ref.~\cite{Bhattacharya:12} to compare with limits from nuclear beta decays. They concluded that the constraints from the radiative pion decays are much more stringent than those from nuclear beta decays.

The analysis of Refs.~\cite{Bhattacharya:12,Cirigliano2013,CiriglianoGardnerHolstein:13} parameterize their interactions in terms of quark-level scalar and tensor effective couplings, $\epsilon_S$ and $\epsilon_T$ coupling to left-handed neutrinos, and $\tilde{\epsilon_S}$ and $\tilde{\epsilon_T}$ coupling to right-handed neutrinos. For the present analysis of the nuclear data we use $C_V=C_V^\prime$ and $C_A=C_A^\prime$ and the two sets of coupling constants are related by
\begin{eqnarray}
g_S \; \epsilon_S = \frac{C_S+C_S^\prime}{2\;C_V}\nonumber\\
g_S \; \tilde{\epsilon_S} = \frac{C_S-C_S^\prime}{2\;C_V}\nonumber\\
g_T \; \epsilon_T = \frac{C_T+C_T^\prime}{8\;C_A}\nonumber\\
g_T \; \tilde{\epsilon_T} = \frac{C_T-C_T^\prime}{8\;C_A}
\label{eq:relations}
\end{eqnarray}
where $g_S$ and $g_T$ are nucleon form factors. These have been calculated using Lattice QCD calculations~\cite{Bhattacharya:12,Green2012,Bhattacharya:13} and are shown in Table~\ref{tab:form factors}.
\begin{table}[!ht]
\caption{Nuclear and pion form factors from calculations. Note that the quoted error bar on the results of Ref.~\cite{Green2012}~and~\cite{Bhattacharya:13} does not include systematic errors~\cite{Lin2013priv}.}
\begin{ruledtabular}
\begin{tabular}{llll}
Ref. & $g^{\rm N}_S$ & $g^{\rm N}_T$   & $f^{\pi}_T$ \\
\hline
\cite{Bhattacharya:12} & $0.8(4)$ & $1.05(35)$ &      \\
\cite{Green2012}& $0.97(12)$ & $1.04(2)$ &      \\
\cite{Bhattacharya:13}& $0.66(24)$ & $1.09(5)$ &      \\
\cite{Mateu2007}  &  &  &     $0.24(4)$ \\
\end{tabular}
\end{ruledtabular}
\label{tab:form factors}
\end{table}
Precision experiments in pion decay observables can also be used to search for tensor currents. The observable in pion radiative decay is sensitive to a product similar to Eq.~\ref{eq:relations} except that there is a pion form factor $f^{\pi }_T$ instead of the nucleon form factor $g^{\rm N}_T$. The pion form factor has been calculated in Ref.~\cite{Mateu2007}. The nucleon form factors were calculated using a renormalization scale of 2 GeV, while Ref.~\cite{Mateu2007} used 1 GeV for the pion. Extending the latter to 2 GeV increases the pion form factor by about 5\%\cite{GGprivate}. Nevertheless, for the analysis presented in this paper, we set $g^{\rm N}_T~\equiv~1$, and $f^{\pi}_T~\equiv~0.24$ and ignore the uncertainties in "translating" pion to nuclear-decay observables. Including the uncertainties decreases the sensitivity of the pion observables when looking at limits on the nuclear-framework couplings we use in this paper.  The uncertainties in the form factors make the nuclear limits worse when looking at the quark couplings because presently the uncertainties in the pion form factor are smaller than those for the nucleon form factors. This shows that work on producing more precise nucleon form factors is important.
%But in order to get an idea on what experiments are more promising to pursue at the present time it is useful to assume that the form factor uncertainties will be reduced by improved calculations.

\section{Data set}

The neutron lifetime $\tau_n$ and $\beta$~asymmetry parameter $A_0$ can be used to determine V$_{ud}$ (e.g.~\cite{Abele2008}~and~\cite{Liu:2010}). However, due to the strong Gamow-Teller character of the neutron decay, the measurements can also be seen as sensitive probes for a tensor-type interaction. We adopt the 2012 Particle Data Group (PDG2012~\cite{pdg:12}) selection; in addition, we include the most recent averages for the UCNA~\cite{Mendenhall2013} and PERKEOII~\cite{Mund2013} experiment (Table~\ref{tab:InputData}). All quoted values for $A_0$ include a $\mathcal{O}$(1~\%) correction for weak magnetism, $g_V~-~g_A$~interference, and nuclear recoil~\cite{Wilkinson1982}, and a $\mathcal{O}$(0.1~\%) radiative correction~\cite{Gluck1992}. In our use of $\tau_n$, we adopt the phase space factor of~\cite{Wilkinson1982} and the overall electro-weak correction factor of~\cite{Marciano2006}. Note that we use the individual data points and their reported errors in our analysis. The inconsistency of the neutron lifetime data is discussed in e.g. Ref.~\cite{pdg:12}~and~\cite{Wietfeldt:2011}. In this paper, the effect of this spread on the obtained limits is illustrated in Fig.~\ref{fig:Neutrondescr}.

For nuclear $\beta$~decays, a limited selection was made of experiments most sensitive to tensor and scalar interactions (Table~\ref{tab:InputData}). The value for the $\beta$-asymmetry parameter, $A_{\beta}$, of $^{60}$Co~\cite{Wauters:2010} includes a recoil correction~\cite{Holstein1974}. The $\beta\nu$-correlation, $a_{\beta\nu}$, results of Ref.~\cite{Adelberger1999} and the value of the $\beta\nu$~correlations for $^6$He of Ref.~\cite{Johnson1963} and for $^{32}$Ar~\cite{Adelberger1999} include the radiative corrections of Ref.~\cite{Gluck:1998}. For the $\beta\nu$~correlation of $^{38m}$K recoil and radiative correction were estimated to be $<$~10$^{-4}$~\cite{Gorelov2005}. A non-zero value of the $b_{Fierz}$ due to scalar weak currents would manifest as a isotope dependent variation in the corrected $ft$-values of the 0$^+$$\rightarrow$0$^+$ super-allowed Fermi transitions. The extraction of $b_{Fierz}$ from the $\mathcal{F}t$ values from these transitions is described in Ref.~\cite{Hardy2005}. We choose not to include the result of ref.~\cite{Pitcairn2009} because their bounds are dominated by the limited knowledge of the recoil-order corrections.

Experiments on radiative pion decays (Bolotov~et~al.~\cite{Bolotov:1990} and PIBETA~Ref.~\cite{Frlez:2004}) have observed discrepancies with the expected spectra which could be interpreted as possible evidence for tensor currents~\cite{Poblaguev:1990}. However, this was contested by Ref.~\cite{Quin1993}. Moreover, more recently the PIBETA collaboration published more precise data with a wider kinematic coverage and showed good agreement with the Standard Model expectations~\cite{bychkov2009}. The upper limit from this last measurement are shown in Fig.~\ref{fig:NuclearOnly},~\ref{fig:All_PERKEOUCNA},~and~\ref{fig:All_PERKEOUCNACSCT}.

\begin{table*}[!htbp]
\setlength{\extrarowheight}{4pt}
\caption{\label{tab:InputData}Selected correlation measurements sensitive to tensor or scalar type weak currents. For the neutron decay data, we include the more recent values for the corrected $\beta$-asymmetry parameter $A_0$ of~\cite{Mund2013} and~\cite{Mendenhall2013} in addition to the Particle Data Group~\cite{pdg:12} selection. The SM value for A$_0$ depends on $\lambda$, the ratio of the axial-vector to vector coupling constants $C_A/C_V$: A$_{0,SM} = \frac{-2(\lambda^2 - |\lambda|)}{1-3\lambda^2}$. The SM value for the neutron lifetime depends on $\lambda$ and V$_{ud}$. V$_{ud}$ is fixed by the corrected ft-values from the 0$^+$$\rightarrow$0$^+$ super-allowed Fermi transitions.}
\begin{ruledtabular}
\begin{tabular}{cccccccc}
Isotope &   Parameter   &   Decay type  &   SM value (q$^2 \rightarrow $0)     &   $\langle\frac{m}{E}\rangle$   &   Value   &   Error   &   Reference \vspace{2pt}\\
\hline

$^6$He              &   $a_{\beta\nu}$         &   $\beta^-$, GT   &   -$\frac{1}{3}$  &   0.286   &   -0.3308 &   0.003   &   \cite{Johnson1963}\cite{Gluck:1998}\\
$^{14}$O~$^{10}$C	& $P_F/P_{GT}$  &   $\beta^+$, F/GT &   1   &   0.292	&   0.9996  &	0.0037	&   \cite{Carnoy:1991} \\
$^{26m}$Al~$^{30}$P & $P_F/P_{GT}$  &   $\beta^+$, F/GT &   1   &  0.216  &   1.003   &   0.0184  &   \cite{Wichers:1987} \\
$^{32}$Ar           &   $a_{\beta\nu}$         &   $\beta^+$, F    &   1   &   0.191  &   0.9989  &   0.0065	&    \cite{Adelberger1999} \\
$^{38m}$K           &   $a_{\beta\nu}$         &   $\beta^+$, F    &   1   &   0.133  &	0.9981  &	0.0045  &   \cite{Gorelov2005}  \\
$^{60}$Co           &   $A_{\beta}$         &   $\beta^-$, GT   &   -1  &   0.704   &   -1.027  &   0.022   &   \cite{Wauters:2010} \\
%$^{114}$In          &   $A$         &   $\beta^-$, GT   &  -1  &   0.209   &   -0.994  &   0.014   &   \cite{Wauters2009c} \\
0$^+$$\rightarrow$0$^+$   &	$b_{Fierz}$   &   $\beta^+$, F    &   0   &   n/a     &   -0.0022 & 	0.0026	&   \cite{Hardy2009}    \vspace{6pt}\\

n                   &   $A_0$         &   $\beta^-$, F/GT & A$_{0,SM}$   &   0.560   &   -0.11952&   0.00110  &   \cite{Mendenhall2013}\cite{Liu:2010}\cite{Plaster:2012}
\\
n                   &   $A_0$         &   $\beta^-$, F/GT & A$_{0,SM}$    &   0.539  &   -0.11926&   0.00050  &   \cite{Mund2013}\cite{Abele2002}\cite{Abele1997}\\
n                   &   $A_0$         &   $\beta^-$, F/GT & A$_{0,SM}$     &  0.582   &   -0.1160 &   0.0015  &   \cite{Liaud1997}\\
n                   &   $A_0$         &   $\beta^-$, F/GT & A$_{0,SM}$    &   0.558   &   -0.1135 &   0.0014  &   \cite{Yerozolimsky1997}\cite{Erozolimsky:1990}\\
n                   &   $A_0$         &   $\beta^-$, F/GT & A$_{0,SM}$     &  0.551   &   -0.1146 &   0.0019  &   \cite{Bopp1986}\vspace{6pt}\\

n                   &   $\tau$      &   $\beta^-$, F/GT &  &  0.653    &   881.6   &   2.1  & \cite{Arzumanov:2012}\cite{Arzumanov2000}
\\
n                   &   $\tau$      &   $\beta^-$, F/GT &  &   0.653    &   880.7   &   1.8     &   \cite{Pichlmaier:2010}
\\
n                   &   $\tau$      &   $\beta^-$, F/GT &  &  0.653    &   886.3   &   3.4     &   \cite{NICO2005}\\
n                   &   $\tau$      &   $\beta^-$, F/GT &  &  0.653    &   878.5   &   0.76    &   \cite{Serebrov2005}\\
n                   &   $\tau$      &   $\beta^-$, F/GT &  &  0.653    &   889.2   &   4.8     &   \cite{Byrne:1996}\\
n                   &   $\tau$      &   $\beta^-$, F/GT &  &  0.653    &   882.6   &   2.7     &   \cite{Mampe:1993}\\
n                   &   $\tau$      &   $\beta^-$, F/GT &  &  0.653    &   887.6   &   3.0     &   \cite{Mampe:1989}\\

\end{tabular}
\end{ruledtabular}
\end{table*}

\section{Method}

Using the expressions from~\cite{Jackson1957a}, a general $\chi^2$ function is constructed in terms of $C_V$, $C_A$, $C_S$, $C^\prime_S$, $C_T$, and $C^\prime_T$. We fix $C_V$ using the corrected $\log{ft}$'s from the $0^+ \rightarrow 0^+$ transitions~\cite{Hardy2009}. A 2-D $\chi^2$ surface is constructed by stepping through different values of the two coupling constants of interest, letting the others vary to minimize it. For example, for the limits on $C_T$-$C^\prime_T$, these are held fix at each point in the $C_T$-$C^\prime_T$ plane while $C_A$, $C_S$, $C^\prime_S$ are varied at each point to minimize $\chi^2$.
%The minimization is performed with the ROOT Minimizer using the Minuit algorithm.
For each point the probability density function is computed as $e^{-\chi^2/2}/N$, where $N$ is a normalization so the sum of all probabilities yields unity. The confidence level contours were obtained as the loci of constant $\chi^2$ that have a probability such that the sum of all points with higher probabilities, i.e. smaller $\chi^2$, yield the desired probability (68\%, 90\%, and 95\% for the plots shown in this paper). The 1-D confidence intervals were calculated from the projected probability density surfaces.

\section{Results}

\begin{figure}[tb]
%\begin{minipage}[b]{0.45\linewidth}
\centering
\includegraphics[width=0.45\textwidth]{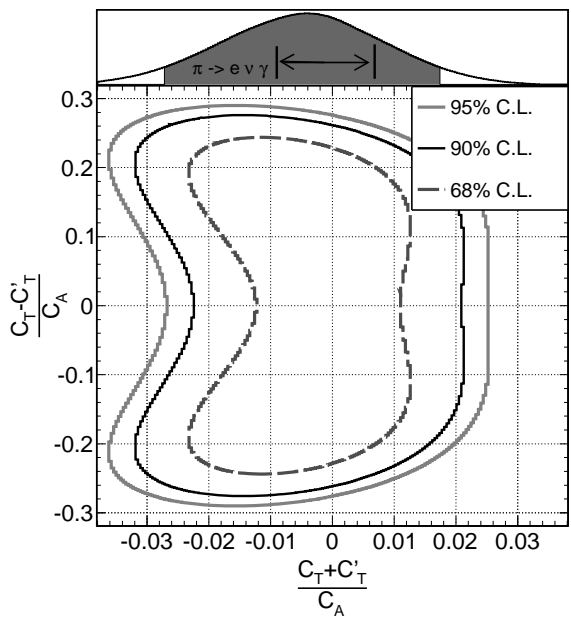}
\caption{68~\%, 90~\% and 95~\% C.L. contours of $C_T$ and $C^\prime_T$ from selected nuclear $\beta$-decay data~(Table~\ref{tab:InputData}) The top panel shows the 1-D projection of the probability distribution, the shaded area is the 90~\% confidence interval for $\frac{C_{T}+C^{'}_T}{C_A}$. }
\label{fig:NuclearOnly}
%\end{minipage}
\end{figure}

\begin{figure}[bt]
%\begin{minipage}[b]{0.45\linewidth}
\centering
\includegraphics[width=0.45\textwidth]{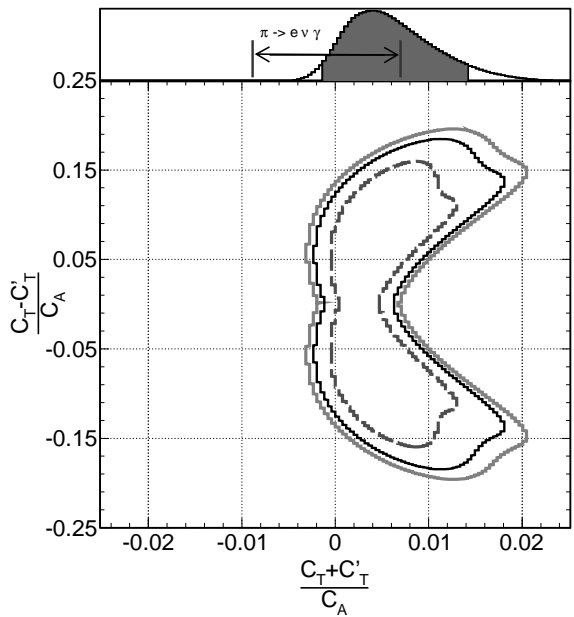}
\caption{Same as Fig.~\protect\ref{fig:NuclearOnly} but including neutron decay data.}
\label{fig:All_PERKEOUCNA}
%\end{minipage}
\end{figure}

Fig.~\ref{fig:NuclearOnly} shows confidence-level contours for $C_T$ and $C^\prime_T$ from nuclear $\beta$ decays excluding neutron decay data. These \emph{nuclear-only} limits are dominated by the $a_{\beta \nu}$ of $^6{\rm He}$ and the relative polarization measurements of Ref.~\cite{Carnoy:1991}. Combining this data with the neutron lifetime and $\beta$ asymmetry further tightens the limits for a tensor interaction coupled to left-handed neutrinos as shown in Fig.~\ref{fig:All_PERKEOUCNA}.
In order to compare with data from radiative pion decay~\cite{bychkov2009} we project the limits from nuclear decay onto the $C_T+C^\prime_T$ axis. As can be seen the nuclear decay limits are comparable to those from pion decays. The 90~\% C.L. intervals on the 1-D projections are $ -0.14 \times 10^{-2} < (C_T+C^\prime_T)/C_A < 1.4 \times 10^{-2} $ and $ -0.16 < (C_T-C^{'}_T)/C_A < 0.16$.

%\end{minipage}
%\hspace{0.5cm}
%\begin{minipage}[b]{0.45\linewidth}

\begin{figure}[htb]
\includegraphics[width=0.45\textwidth]{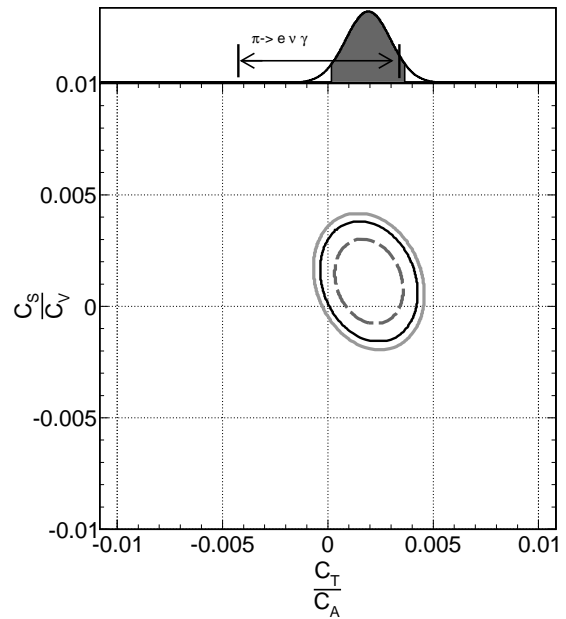}
\caption{Limits $C_T$ and $C_S$ combining neutron and nuclear $\beta$ decay data for the 3-parameter fit. On top we show the probability distribution of the limits on $C_T$ obtained by projecting the 2D distribution and compare to the limits from pion decay data.}
\label{fig:All_PERKEOUCNACSCT}
\end{figure}

Fig.~\ref{fig:All_PERKEOUCNACSCT} shows limits for the tensor and scalar currents assuming they couple only to left-handed neutrinos~($C^\prime_T = C_T$ and $C^\prime_S = C_S$), with the limits on $C_S$ dominated by the $b_{\rm Fierz}$ $0^+ \rightarrow 0^+$ super-allowed transitions. The corresponding 90\%~C.L. for $C_S$ and $C_T $ are
$-0.1 \times 10^{-2} < C_S/C_V < 0.3 \times 10^{-2}$ and $-0.0 \times 10^{-2} < C_T/C_A < 0.4 \times 10^{-2}$.
For this 3-parameter fit, the limits on tensor currents from $\beta$-decays are more stringent than those from pion decays and are in agreement with the evaluation of ref.~\cite{Pattie2013}.
The value of $\lambda$ (~=~C$_A$/C$_V$~) at the minimum $\chi^2$, $-1.2753(6)$, is consistent with the PDG2012 recommended value of $-1.2701(25)$.

\begin{figure}[ht]
%\begin{minipage}[b]{0.45\linewidth}
\centering
\includegraphics[width=0.45\textwidth]{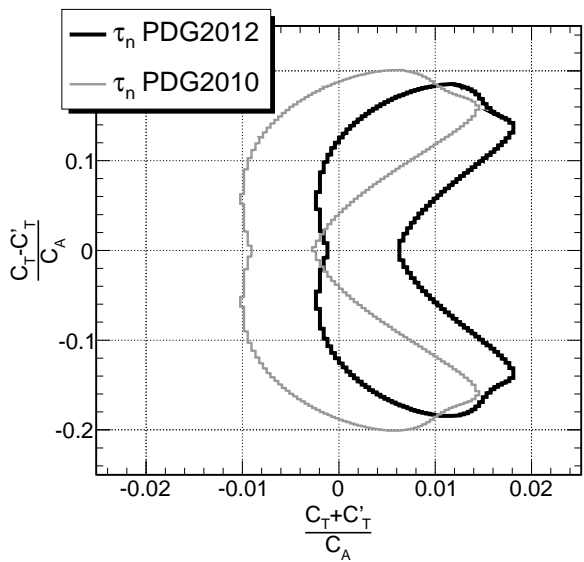}
\caption{The effect of the 5.6~s shift between the PDG2010 or PDG2012 recommended values for the neutron lifetimes on the 90\% confidence limits on C$_T$ and C$^\prime_T$.}
\label{fig:Neutrondescr}
%\end{minipage}
\end{figure}
%\hspace{0.5cm}
%\begin{minipage}[b]{0.45\linewidth}
\begin{figure}
\centering
\includegraphics[width=0.45\textwidth]{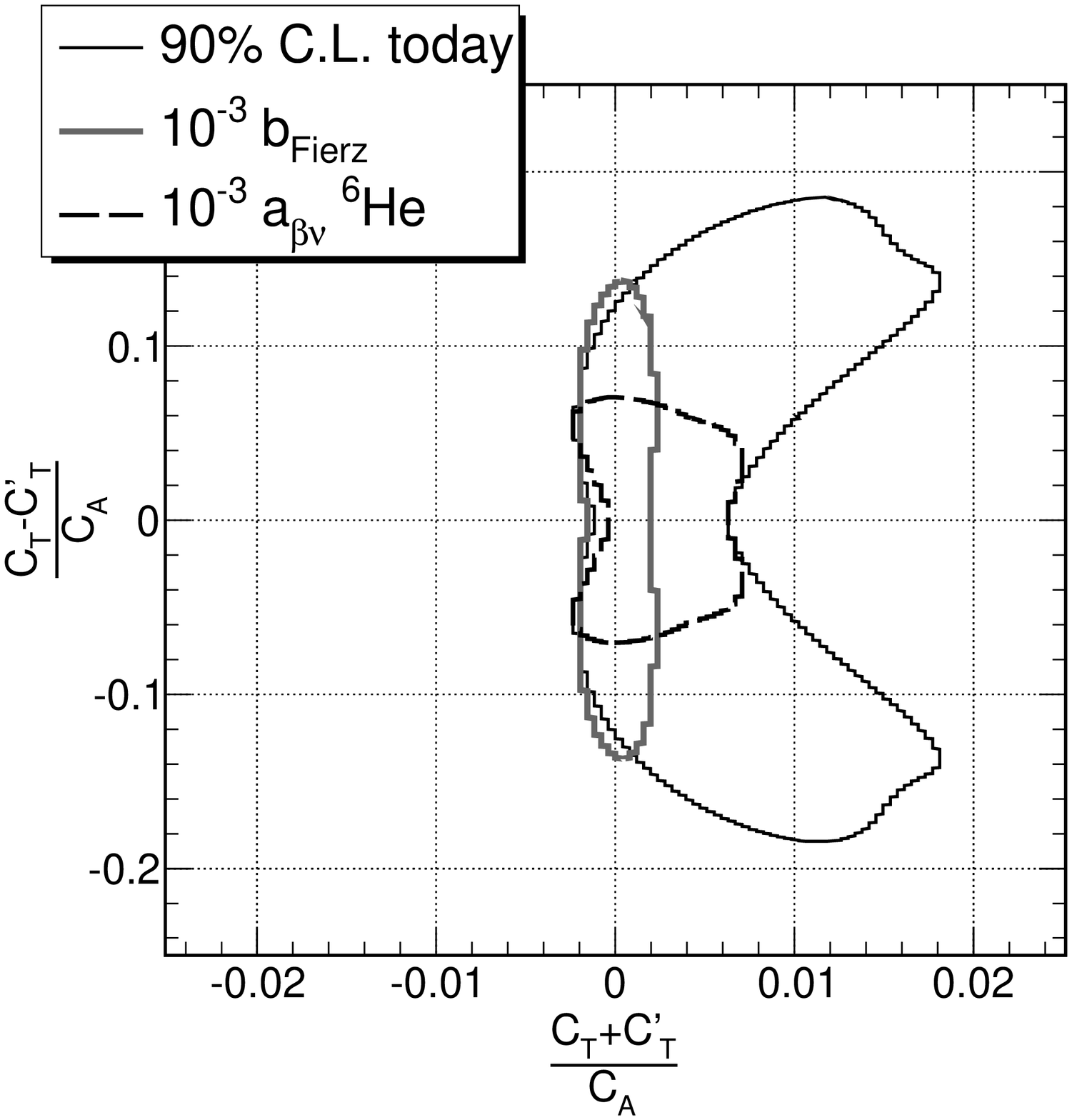}
\caption{Limits on tensor currents from envisaged measurements of the correlation coefficient for $^6$He at $0.1$\%~\cite{Knecht2011} or of the Fierz interference term $b_{Fierz}$ to $10^{-3}$ in $^6$He or neutron decays~\cite{Pocanic2009}. In the formalism of Eq.~\ref{eqn:jtw}, the current LHC limits from Ref.~\cite{Cirigliano2013}  are $|(C_T+C^{'}_{T})/C_A|< 6 \times 10^{-3}$ and $|(C_T-C^{'}_{T})/C_A|< 2 \times 10^{-2}$.}
\label{fig:Future}
%\end{minipage}
\end{figure}

Nuclear $\beta$ decay data from Table~\ref{tab:InputData} {\em without} neutron-decay information form a consistent data set ($p$-value of $0.4$ for Fig.~\ref{fig:NuclearOnly}). In contrast, there is a large spread in the neutron decay data~(Tab.~\ref{tab:InputData}). Fig.~\ref{fig:Neutrondescr} illustrates the sensitivity of the obtained limits to inconsistencies in neutron data by adopting the two different values for $\tau_n$, the 2010 and 2012 PDG recommended values, which differ by 5~$\sigma$. Note that including the most recent neutron data, with A$_0$ dominated by the the PERKEOII and UCNA experiments and the PDG2012 $\tau_n$, significantly tightens the limits on tensor currents coupled to left-handed neutrinos (Fig.~\ref{fig:Neutrondescr}). This also manifests in more than two times tighter limits on $C_T/C_A$ compared to the previous evaluation Ref.~\cite{Severijns2006} in the 3-parameter fit.

\section{Conclusions}

Present $\beta$-decay data is consistent with $C_S = C^\prime_S = C_T = C^\prime_T=0$. This conclusion is in agreement with previous evaluations (e.g. Boothroyd et al.~\cite{Boothroyd:1984}~and~Severijns et al.~\cite{Severijns2006}) and is independent of possible inconsistencies in the neutron decay lifetime or in the asymmetry parameter.

The limits from nuclear and neutron $\beta$ decays are more stringent than those from pion decays. As shown in Fig.~\ref{fig:Future}, a future correlation-parameter measurement with an uncertainty of $10^{-3}$ in neutron or selected nuclear decays where the higher-order corrections are under control, would significantly improve the discovery potential. This would surpass the sensitivity of the current and future LHC experiments for new physics emerging at a higher energy scale than the production threshold~\cite{Cirigliano2013,Gonzalez2013}. In addition, a reduced theoretical uncertainty of the nucleon and pion form factors would enable a combined analysis of all low energy data to further constrain tensor-type weak currents.

% Once the uncertainty on g$_T$ and the corresponding f$_T$
%I see the problem.  Indeed, even the radiative pion decay involves its own matrix element
%(analogue of gT). So multiple uncertainties (=dilutions) arise.
%Perhaps you could mention that a combined analysis will increase the constraining power,
%once the uncertainty on gT  gets under control (let's say, at the 10% level?).

\section*{Acknowledgments}
We thank Oscar Naviliat-Cuncic, Peter M\"{u}ller, Martin Gonz\'{a}lez-Alonso and Vincenzo Cirigliano for their fruitful comments.
This work was supported by the US DOE under Contract No. DE-FG02-97ER41020.

%\bibliography{CTLimitsReSubm.bbl}
\bibliography{bibliography_database.bib}

\begin{thebibliography}{65}%
\makeatletter
\providecommand \@ifxundefined [1]{%
 \@ifx{#1\undefined}
}%
\providecommand \@ifnum [1]{%
 \ifnum #1\expandafter \@firstoftwo
 \else \expandafter \@secondoftwo
 \fi
}%
\providecommand \@ifx [1]{%
 \ifx #1\expandafter \@firstoftwo
 \else \expandafter \@secondoftwo
 \fi
}%
\providecommand \natexlab [1]{#1}%
\providecommand \enquote  [1]{``#1''}%
\providecommand \bibnamefont  [1]{#1}%
\providecommand \bibfnamefont [1]{#1}%
\providecommand \citenamefont [1]{#1}%
\providecommand \href@noop [0]{\@secondoftwo}%
\providecommand \href [0]{\begingroup \@sanitize@url \@href}%
\providecommand \@href[1]{\@@startlink{#1}\@@href}%
\providecommand \@@href[1]{\endgroup#1\@@endlink}%
\providecommand \@sanitize@url [0]{\catcode `\\12\catcode `\$12\catcode
  `\&12\catcode `\#12\catcode `\^12\catcode `\_12\catcode `\%12\relax}%
\providecommand \@@startlink[1]{}%
\providecommand \@@endlink[0]{}%
\providecommand \url  [0]{\begingroup\@sanitize@url \@url }%
\providecommand \@url [1]{\endgroup\@href {#1}{\urlprefix }}%
\providecommand \urlprefix  [0]{URL }%
\providecommand \Eprint [0]{\href }%
\providecommand \doibase [0]{http://dx.doi.org/}%
\providecommand \selectlanguage [0]{\@gobble}%
\providecommand \bibinfo  [0]{\@secondoftwo}%
\providecommand \bibfield  [0]{\@secondoftwo}%
\providecommand \translation [1]{[#1]}%
\providecommand \BibitemOpen [0]{}%
\providecommand \bibitemStop [0]{}%
\providecommand \bibitemNoStop [0]{.\EOS\space}%
\providecommand \EOS [0]{\spacefactor3000\relax}%
\providecommand \BibitemShut  [1]{\csname bibitem#1\endcsname}%
\let\auto@bib@innerbib\@empty
%</preamble>
\bibitem [{\citenamefont {Buchm\"uller}\ \emph {et~al.}(1987)\citenamefont
  {Buchm\"uller}, \citenamefont {R\"uckl},\ and\ \citenamefont
  {Wyler}}]{Buchmuller1987442}%
  \BibitemOpen
  \bibfield  {author} {\bibinfo {author} {\bibfnamefont {W.}~\bibnamefont
  {Buchm\"uller}}, \bibinfo {author} {\bibfnamefont {R.}~\bibnamefont
  {R\"uckl}}, \ and\ \bibinfo {author} {\bibfnamefont {D.}~\bibnamefont
  {Wyler}},\ }\href {\doibase 10.1016/0370-2693(87)90637-X} {\bibfield
  {journal} {\bibinfo  {journal} {Physics Letters B}\ }\textbf {\bibinfo
  {volume} {191}},\ \bibinfo {pages} {442 } (\bibinfo {year}
  {1987})}\BibitemShut {NoStop}%
\bibitem [{\citenamefont {Herczeg}(2001)}]{Herczeg2001}%
  \BibitemOpen
  \bibfield  {author} {\bibinfo {author} {\bibfnamefont {P.}~\bibnamefont
  {Herczeg}},\ }\href@noop {} {\bibfield  {journal} {\bibinfo  {journal}
  {Progress in Particle and Nuclear Physics}\ }\textbf {\bibinfo {volume}
  {46}},\ \bibinfo {pages} {413 } (\bibinfo {year} {2001})}\BibitemShut
  {NoStop}%
\bibitem [{\citenamefont {Profumo}\ \emph {et~al.}(2007)\citenamefont
  {Profumo}, \citenamefont {Ramsey-Musolf},\ and\ \citenamefont
  {Tulin}}]{profumo2007}%
  \BibitemOpen
  \bibfield  {author} {\bibinfo {author} {\bibfnamefont {S.}~\bibnamefont
  {Profumo}}, \bibinfo {author} {\bibfnamefont {M.~J.}\ \bibnamefont
  {Ramsey-Musolf}}, \ and\ \bibinfo {author} {\bibfnamefont {S.}~\bibnamefont
  {Tulin}},\ }\href {\doibase 10.1103/PhysRevD.75.075017} {\bibfield  {journal}
  {\bibinfo  {journal} {Phys. Rev. D}\ }\textbf {\bibinfo {volume} {75}},\
  \bibinfo {pages} {075017} (\bibinfo {year} {2007})}\BibitemShut {NoStop}%
\bibitem [{\citenamefont {Belyaev}\ and\ \citenamefont
  {Kogan}(1992)}]{Belyaev1992}%
  \BibitemOpen
  \bibfield  {author} {\bibinfo {author} {\bibfnamefont {V.}~\bibnamefont
  {Belyaev}}\ and\ \bibinfo {author} {\bibfnamefont {I.}~\bibnamefont
  {Kogan}},\ }\href@noop {} {\bibfield  {journal} {\bibinfo  {journal} {Physics
  Letters B}\ }\textbf {\bibinfo {volume} {280}},\ \bibinfo {pages} {238 }
  (\bibinfo {year} {1992})}\BibitemShut {NoStop}%
\bibitem [{\citenamefont {Severijns}\ \emph {et~al.}(2006)\citenamefont
  {Severijns}, \citenamefont {Beck},\ and\ \citenamefont
  {Naviliat-Cuncic}}]{Severijns2006}%
  \BibitemOpen
  \bibfield  {author} {\bibinfo {author} {\bibfnamefont {N.}~\bibnamefont
  {Severijns}}, \bibinfo {author} {\bibfnamefont {M.}~\bibnamefont {Beck}}, \
  and\ \bibinfo {author} {\bibfnamefont {O.}~\bibnamefont {Naviliat-Cuncic}},\
  }\href@noop {} {\bibfield  {journal} {\bibinfo  {journal} {Reviews of Modern
  Physics}\ }\textbf {\bibinfo {volume} {78}},\ \bibinfo {pages} {991}
  (\bibinfo {year} {2006})}\BibitemShut {NoStop}%
\bibitem [{\citenamefont {Behr}\ and\ \citenamefont
  {Gwinner}(2009)}]{Behr2009}%
  \BibitemOpen
  \bibfield  {author} {\bibinfo {author} {\bibfnamefont {J.~A.}\ \bibnamefont
  {Behr}}\ and\ \bibinfo {author} {\bibfnamefont {G.}~\bibnamefont {Gwinner}},\
  }\href@noop {} {\bibfield  {journal} {\bibinfo  {journal} {Journal of Physics
  G}\ }\textbf {\bibinfo {volume} {36}},\ \bibinfo {pages} {033101 (38pp)}
  (\bibinfo {year} {2009})}\BibitemShut {NoStop}%
\bibitem [{\citenamefont {Abele}(2008)}]{Abele2008}%
  \BibitemOpen
  \bibfield  {author} {\bibinfo {author} {\bibfnamefont {H.}~\bibnamefont
  {Abele}},\ }\href {\doibase 10.1016/j.ppnp.2007.05.002} {\bibfield  {journal}
  {\bibinfo  {journal} {Progress in Particle and Nuclear Physics}\ }\textbf
  {\bibinfo {volume} {60}},\ \bibinfo {pages} {1 } (\bibinfo {year}
  {2008})}\BibitemShut {NoStop}%
\bibitem [{\citenamefont {Dubbers}\ and\ \citenamefont
  {Schmidt}(2011)}]{Dubbers:2011}%
  \BibitemOpen
  \bibfield  {author} {\bibinfo {author} {\bibfnamefont {D.}~\bibnamefont
  {Dubbers}}\ and\ \bibinfo {author} {\bibfnamefont {M.~G.}\ \bibnamefont
  {Schmidt}},\ }\href@noop {} {\bibfield  {journal} {\bibinfo  {journal} {Rev.
  Mod. Phys.}\ }\textbf {\bibinfo {volume} {83}},\ \bibinfo {pages} {1111}
  (\bibinfo {year} {2011})}\BibitemShut {NoStop}%
\bibitem [{\citenamefont {Po\v{c}ani\'c}(2010)}]{Pocanic201026c}%
  \BibitemOpen
  \bibfield  {author} {\bibinfo {author} {\bibfnamefont {D.}~\bibnamefont
  {Po\v{c}ani\'c}},\ }\href@noop {} {\bibfield  {journal} {\bibinfo  {journal}
  {Nuclear Physics A}\ }\textbf {\bibinfo {volume} {844}},\ \bibinfo {pages}
  {26c } (\bibinfo {year} {2010})}\BibitemShut {NoStop}%
\bibitem [{\citenamefont {Ito}\ and\ \citenamefont
  {Pr\'ezeau}(2005)}]{Ito2005}%
  \BibitemOpen
  \bibfield  {author} {\bibinfo {author} {\bibfnamefont {T.~M.}\ \bibnamefont
  {Ito}}\ and\ \bibinfo {author} {\bibfnamefont {G.}~\bibnamefont
  {Pr\'ezeau}},\ }\href@noop {} {\bibfield  {journal} {\bibinfo  {journal}
  {Phys. Rev. Lett.}\ }\textbf {\bibinfo {volume} {94}},\ \bibinfo {pages}
  {161802} (\bibinfo {year} {2005})}\BibitemShut {NoStop}%
\bibitem [{\citenamefont {Bhattacharya}\ \emph {et~al.}(2012)\citenamefont
  {Bhattacharya}, \citenamefont {Cirigliano}, \citenamefont {Cohen},
  \citenamefont {Filipuzzi}, \citenamefont {Gonz\'alez-Alonso}, \citenamefont
  {Graesser}, \citenamefont {Gupta},\ and\ \citenamefont
  {Lin}}]{Bhattacharya:12}%
  \BibitemOpen
  \bibfield  {author} {\bibinfo {author} {\bibfnamefont {T.}~\bibnamefont
  {Bhattacharya}}, \bibinfo {author} {\bibfnamefont {V.}~\bibnamefont
  {Cirigliano}}, \bibinfo {author} {\bibfnamefont {S.~D.}\ \bibnamefont
  {Cohen}}, \bibinfo {author} {\bibfnamefont {A.}~\bibnamefont {Filipuzzi}},
  \bibinfo {author} {\bibfnamefont {M.}~\bibnamefont {Gonz\'alez-Alonso}},
  \bibinfo {author} {\bibfnamefont {M.~L.}\ \bibnamefont {Graesser}}, \bibinfo
  {author} {\bibfnamefont {R.}~\bibnamefont {Gupta}}, \ and\ \bibinfo {author}
  {\bibfnamefont {H.-W.}\ \bibnamefont {Lin}},\ }\href {\doibase
  10.1103/PhysRevD.85.054512} {\bibfield  {journal} {\bibinfo  {journal} {Phys.
  Rev. D}\ }\textbf {\bibinfo {volume} {85}},\ \bibinfo {pages} {054512}
  (\bibinfo {year} {2012})}\BibitemShut {NoStop}%
\bibitem [{\citenamefont {Cirigliano}\ \emph {et~al.}()\citenamefont
  {Cirigliano}, \citenamefont {Gonzalez-Alonso},\ and\ \citenamefont
  {Graesser}}]{Cirigliano2013}%
  \BibitemOpen
  \bibfield  {author} {\bibinfo {author} {\bibfnamefont {V.}~\bibnamefont
  {Cirigliano}}, \bibinfo {author} {\bibfnamefont {M.}~\bibnamefont
  {Gonzalez-Alonso}}, \ and\ \bibinfo {author} {\bibfnamefont {M.~L.}\
  \bibnamefont {Graesser}},\ }\href@noop {} {\bibfield  {journal} {\bibinfo
  {journal} {Journal Of High Energy Physics}\ }\textbf {\bibinfo {volume}
  {1302}}}\BibitemShut {NoStop}%
\bibitem [{\citenamefont {Cirigliano}\ \emph {et~al.}(2013)\citenamefont
  {Cirigliano}, \citenamefont {Gardner},\ and\ \citenamefont
  {Holstein}}]{CiriglianoGardnerHolstein:13}%
  \BibitemOpen
  \bibfield  {author} {\bibinfo {author} {\bibfnamefont {V.}~\bibnamefont
  {Cirigliano}}, \bibinfo {author} {\bibfnamefont {S.}~\bibnamefont {Gardner}},
  \ and\ \bibinfo {author} {\bibfnamefont {B.~R.}\ \bibnamefont {Holstein}},\
  }\href@noop {} {\bibfield  {journal} {\bibinfo  {journal} {Progress in
  Particle and Nucl. Phys.}\ }\textbf {\bibinfo {volume} {71}},\ \bibinfo
  {pages} {93} (\bibinfo {year} {2013})}\BibitemShut {NoStop}%
\bibitem [{\citenamefont {Bychkov}\ \emph {et~al.}(2009)\citenamefont
  {Bychkov}, \citenamefont {Pocanic}, \citenamefont {VanDevender},
  \citenamefont {Baranov}, \citenamefont {Bertl}, \citenamefont {Bystritsky},
  \citenamefont {Frlez}, \citenamefont {Kalinnikov}, \citenamefont {Khomutov},
  \citenamefont {Korenchenko}, \citenamefont {Korenchenko}, \citenamefont
  {Korolija}, \citenamefont {Kozlowski}, \citenamefont {Kravchuk},
  \citenamefont {Kuchinsky}, \citenamefont {Li}, \citenamefont {Mekterovic},
  \citenamefont {Mzhavia}, \citenamefont {Ritt}, \citenamefont {Robmann},
  \citenamefont {Rondon-Aramayo}, \citenamefont {Rozhdestvensky}, \citenamefont
  {Sakhelashvili}, \citenamefont {Scheu}, \citenamefont {Straumann},
  \citenamefont {Supek}, \citenamefont {Tsamalaidze}, \citenamefont {van~der
  Schaaf}, \citenamefont {Velicheva}, \citenamefont {Volnykh}, \citenamefont
  {Wang},\ and\ \citenamefont {Wirtz}}]{bychkov2009}%
  \BibitemOpen
  \bibfield  {author} {\bibinfo {author} {\bibfnamefont {M.}~\bibnamefont
  {Bychkov}}, \bibinfo {author} {\bibfnamefont {D.}~\bibnamefont {Pocanic}},
  \bibinfo {author} {\bibfnamefont {B.~A.}\ \bibnamefont {VanDevender}},
  \bibinfo {author} {\bibfnamefont {V.~A.}\ \bibnamefont {Baranov}}, \bibinfo
  {author} {\bibfnamefont {W.}~\bibnamefont {Bertl}}, \bibinfo {author}
  {\bibfnamefont {Y.~M.}\ \bibnamefont {Bystritsky}}, \bibinfo {author}
  {\bibfnamefont {E.}~\bibnamefont {Frlez}}, \bibinfo {author} {\bibfnamefont
  {V.~A.}\ \bibnamefont {Kalinnikov}}, \bibinfo {author} {\bibfnamefont
  {N.~V.}\ \bibnamefont {Khomutov}}, \bibinfo {author} {\bibfnamefont {A.~S.}\
  \bibnamefont {Korenchenko}}, \bibinfo {author} {\bibfnamefont {S.~M.}\
  \bibnamefont {Korenchenko}}, \bibinfo {author} {\bibfnamefont
  {M.}~\bibnamefont {Korolija}}, \bibinfo {author} {\bibfnamefont
  {T.}~\bibnamefont {Kozlowski}}, \bibinfo {author} {\bibfnamefont {N.~P.}\
  \bibnamefont {Kravchuk}}, \bibinfo {author} {\bibfnamefont {N.~A.}\
  \bibnamefont {Kuchinsky}}, \bibinfo {author} {\bibfnamefont {W.}~\bibnamefont
  {Li}}, \bibinfo {author} {\bibfnamefont {D.}~\bibnamefont {Mekterovic}},
  \bibinfo {author} {\bibfnamefont {D.}~\bibnamefont {Mzhavia}}, \bibinfo
  {author} {\bibfnamefont {S.}~\bibnamefont {Ritt}}, \bibinfo {author}
  {\bibfnamefont {P.}~\bibnamefont {Robmann}}, \bibinfo {author} {\bibfnamefont
  {O.~A.}\ \bibnamefont {Rondon-Aramayo}}, \bibinfo {author} {\bibfnamefont
  {A.~M.}\ \bibnamefont {Rozhdestvensky}}, \bibinfo {author} {\bibfnamefont
  {T.}~\bibnamefont {Sakhelashvili}}, \bibinfo {author} {\bibfnamefont
  {S.}~\bibnamefont {Scheu}}, \bibinfo {author} {\bibfnamefont
  {U.}~\bibnamefont {Straumann}}, \bibinfo {author} {\bibfnamefont
  {I.}~\bibnamefont {Supek}}, \bibinfo {author} {\bibfnamefont
  {Z.}~\bibnamefont {Tsamalaidze}}, \bibinfo {author} {\bibfnamefont
  {A.}~\bibnamefont {van~der Schaaf}}, \bibinfo {author} {\bibfnamefont
  {E.~P.}\ \bibnamefont {Velicheva}}, \bibinfo {author} {\bibfnamefont {V.~P.}\
  \bibnamefont {Volnykh}}, \bibinfo {author} {\bibfnamefont {Y.}~\bibnamefont
  {Wang}}, \ and\ \bibinfo {author} {\bibfnamefont {H.-P.}\ \bibnamefont
  {Wirtz}},\ }\href {\doibase 10.1103/PhysRevLett.103.051802} {\bibfield
  {journal} {\bibinfo  {journal} {Phys. Rev. Lett.}\ }\textbf {\bibinfo
  {volume} {103}},\ \bibinfo {pages} {051802} (\bibinfo {year}
  {2009})}\BibitemShut {NoStop}%
\bibitem [{\citenamefont {Pattie}\ \emph {et~al.}(2013)\citenamefont {Pattie},
  \citenamefont {Hickerson},\ and\ \citenamefont {Young}}]{Pattie2013}%
  \BibitemOpen
  \bibfield  {author} {\bibinfo {author} {\bibfnamefont {R.~W.}\ \bibnamefont
  {Pattie}}, \bibinfo {author} {\bibfnamefont {K.~P.}\ \bibnamefont
  {Hickerson}}, \ and\ \bibinfo {author} {\bibfnamefont {A.~R.}\ \bibnamefont
  {Young}},\ }\href@noop {} {\bibfield  {journal} {\bibinfo  {journal} {Phys.
  Rev. C}\ }\textbf {\bibinfo {volume} {88}},\ \bibinfo {pages} {048501}
  (\bibinfo {year} {2013})}\BibitemShut {NoStop}%
\bibitem [{\citenamefont {Jackson}\ \emph
  {et~al.}(1957{\natexlab{a}})\citenamefont {Jackson}, \citenamefont
  {Treiman},\ and\ \citenamefont {Wyld}}]{Jackson1957}%
  \BibitemOpen
  \bibfield  {author} {\bibinfo {author} {\bibfnamefont {J.}~\bibnamefont
  {Jackson}}, \bibinfo {author} {\bibfnamefont {S.}~\bibnamefont {Treiman}}, \
  and\ \bibinfo {author} {\bibfnamefont {J.~H.}\ \bibnamefont {Wyld}},\
  }\href@noop {} {\bibfield  {journal} {\bibinfo  {journal} {Physical Review}\
  }\textbf {\bibinfo {volume} {106}},\ \bibinfo {pages} {517 } (\bibinfo {year}
  {1957}{\natexlab{a}})}\BibitemShut {NoStop}%
\bibitem [{Note1()}]{Note1}%
  \BibitemOpen
  \bibinfo {note} {We neglect pseudo-scalar currents because their role is
  minimal due the slow nature of the nucleon motion within the
  nucleus}\BibitemShut {NoStop}%
\bibitem [{\citenamefont {Jackson}\ \emph
  {et~al.}(1957{\natexlab{b}})\citenamefont {Jackson}, \citenamefont
  {Treiman},\ and\ \citenamefont {Wyld}}]{Jackson1957a}%
  \BibitemOpen
  \bibfield  {author} {\bibinfo {author} {\bibfnamefont {J.}~\bibnamefont
  {Jackson}}, \bibinfo {author} {\bibfnamefont {S.}~\bibnamefont {Treiman}}, \
  and\ \bibinfo {author} {\bibfnamefont {J.~H.}\ \bibnamefont {Wyld}},\
  }\href@noop {} {\bibfield  {journal} {\bibinfo  {journal} {Nuclear Physics}\
  }\textbf {\bibinfo {volume} {4}},\ \bibinfo {pages} {206 } (\bibinfo {year}
  {1957}{\natexlab{b}})}\BibitemShut {NoStop}%
\bibitem [{\citenamefont {Green}\ \emph {et~al.}(2012)\citenamefont {Green},
  \citenamefont {Negele}, \citenamefont {Pochinsky}, \citenamefont {Syritsyn},
  \citenamefont {Engelhardt},\ and\ \citenamefont {Krieg}}]{Green2012}%
  \BibitemOpen
  \bibfield  {author} {\bibinfo {author} {\bibfnamefont {J.~R.}\ \bibnamefont
  {Green}}, \bibinfo {author} {\bibfnamefont {J.~W.}\ \bibnamefont {Negele}},
  \bibinfo {author} {\bibfnamefont {A.~V.}\ \bibnamefont {Pochinsky}}, \bibinfo
  {author} {\bibfnamefont {S.~N.}\ \bibnamefont {Syritsyn}}, \bibinfo {author}
  {\bibfnamefont {M.}~\bibnamefont {Engelhardt}}, \ and\ \bibinfo {author}
  {\bibfnamefont {S.}~\bibnamefont {Krieg}},\ }\href {\doibase
  10.1103/PhysRevD.86.114509} {\bibfield  {journal} {\bibinfo  {journal} {Phys.
  Rev. D}\ }\textbf {\bibinfo {volume} {86}},\ \bibinfo {pages} {114509}
  (\bibinfo {year} {2012})}\BibitemShut {NoStop}%
\bibitem [{\citenamefont {Bhattacharya}\ \emph {et~al.}(2013)\citenamefont
  {Bhattacharya}, \citenamefont {Cohen}, \citenamefont {Gupta}, \citenamefont
  {Joseph},\ and\ \citenamefont {Lin}}]{Bhattacharya:13}%
  \BibitemOpen
  \bibfield  {author} {\bibinfo {author} {\bibfnamefont {T.}~\bibnamefont
  {Bhattacharya}}, \bibinfo {author} {\bibfnamefont {S.~D.}\ \bibnamefont
  {Cohen}}, \bibinfo {author} {\bibfnamefont {R.}~\bibnamefont {Gupta}},
  \bibinfo {author} {\bibfnamefont {A.}~\bibnamefont {Joseph}}, \ and\ \bibinfo
  {author} {\bibfnamefont {H.-W.}\ \bibnamefont {Lin}},\ }\href
  {http://arxiv.org/abs/arXiv:1306.5435} {\bibfield  {journal} {\bibinfo
  {journal} {ArXiv}\ ,\ \bibinfo {pages} {1306.5435}} (\bibinfo {year}
  {2013})}\BibitemShut {NoStop}%
\bibitem [{\citenamefont {Lin}()}]{Lin2013priv}%
  \BibitemOpen
  \bibfield  {author} {\bibinfo {author} {\bibfnamefont {H.-W.}\ \bibnamefont
  {Lin}},\ }\href@noop {} {}\bibinfo {note} {Private communication}\BibitemShut
  {NoStop}%
\bibitem [{\citenamefont {Mateu}\ and\ \citenamefont
  {Portoles}(2007)}]{Mateu2007}%
  \BibitemOpen
  \bibfield  {author} {\bibinfo {author} {\bibfnamefont {V.}~\bibnamefont
  {Mateu}}\ and\ \bibinfo {author} {\bibfnamefont {J.}~\bibnamefont
  {Portoles}},\ }\href@noop {} {\bibfield  {journal} {\bibinfo  {journal}
  {European Physical Journal C}\ }\textbf {\bibinfo {volume} {52}},\ \bibinfo
  {pages} {325} (\bibinfo {year} {2007})}\BibitemShut {NoStop}%
\bibitem [{\citenamefont {Gonz\'alez-Alonso}()}]{GGprivate}%
  \BibitemOpen
  \bibfield  {author} {\bibinfo {author} {\bibfnamefont {M.}~\bibnamefont
  {Gonz\'alez-Alonso}},\ }\href@noop {} {}\bibinfo {note} {Private
  communication}\BibitemShut {NoStop}%
\bibitem [{\citenamefont {Liu}\ \emph {et~al.}(2010)\citenamefont {Liu},
  \citenamefont {Mendenhall}, \citenamefont {Holley}, \citenamefont {Back},
  \citenamefont {Bowles}, \citenamefont {Broussard}, \citenamefont {Carr},
  \citenamefont {Clayton}, \citenamefont {Currie}, \citenamefont {Filippone},
  \citenamefont {Garc\'{i}a}, \citenamefont {Geltenbort}, \citenamefont
  {Hickerson}, \citenamefont {Hoagland}, \citenamefont {Hogan}, \citenamefont
  {Hona}, \citenamefont {Ito}, \citenamefont {Liu}, \citenamefont {Makela},
  \citenamefont {Mammei}, \citenamefont {Martin}, \citenamefont {Melconian},
  \citenamefont {Morris}, \citenamefont {Pattie}, \citenamefont
  {P\'erez~Galv\'an}, \citenamefont {Pitt}, \citenamefont {Plaster},
  \citenamefont {Ramsey}, \citenamefont {Rios}, \citenamefont {Russell},
  \citenamefont {Saunders}, \citenamefont {Seestrom}, \citenamefont {Sondheim},
  \citenamefont {Tatar}, \citenamefont {Vogelaar}, \citenamefont {VornDick},
  \citenamefont {Wrede}, \citenamefont {Yan}, \citenamefont {Young},\ and\
  \citenamefont {Zeck}}]{Liu:2010}%
  \BibitemOpen
  \bibfield  {author} {\bibinfo {author} {\bibfnamefont {J.}~\bibnamefont
  {Liu}}, \bibinfo {author} {\bibfnamefont {M.~P.}\ \bibnamefont {Mendenhall}},
  \bibinfo {author} {\bibfnamefont {A.~T.}\ \bibnamefont {Holley}}, \bibinfo
  {author} {\bibfnamefont {H.~O.}\ \bibnamefont {Back}}, \bibinfo {author}
  {\bibfnamefont {T.~J.}\ \bibnamefont {Bowles}}, \bibinfo {author}
  {\bibfnamefont {L.~J.}\ \bibnamefont {Broussard}}, \bibinfo {author}
  {\bibfnamefont {R.}~\bibnamefont {Carr}}, \bibinfo {author} {\bibfnamefont
  {S.}~\bibnamefont {Clayton}}, \bibinfo {author} {\bibfnamefont
  {S.}~\bibnamefont {Currie}}, \bibinfo {author} {\bibfnamefont {B.~W.}\
  \bibnamefont {Filippone}}, \bibinfo {author} {\bibfnamefont {A.}~\bibnamefont
  {Garc\'{i}a}}, \bibinfo {author} {\bibfnamefont {P.}~\bibnamefont
  {Geltenbort}}, \bibinfo {author} {\bibfnamefont {K.~P.}\ \bibnamefont
  {Hickerson}}, \bibinfo {author} {\bibfnamefont {J.}~\bibnamefont {Hoagland}},
  \bibinfo {author} {\bibfnamefont {G.~E.}\ \bibnamefont {Hogan}}, \bibinfo
  {author} {\bibfnamefont {B.}~\bibnamefont {Hona}}, \bibinfo {author}
  {\bibfnamefont {T.~M.}\ \bibnamefont {Ito}}, \bibinfo {author} {\bibfnamefont
  {C.-Y.}\ \bibnamefont {Liu}}, \bibinfo {author} {\bibfnamefont
  {M.}~\bibnamefont {Makela}}, \bibinfo {author} {\bibfnamefont {R.~R.}\
  \bibnamefont {Mammei}}, \bibinfo {author} {\bibfnamefont {J.~W.}\
  \bibnamefont {Martin}}, \bibinfo {author} {\bibfnamefont {D.}~\bibnamefont
  {Melconian}}, \bibinfo {author} {\bibfnamefont {C.~L.}\ \bibnamefont
  {Morris}}, \bibinfo {author} {\bibfnamefont {R.~W.}\ \bibnamefont {Pattie}},
  \bibinfo {author} {\bibfnamefont {A.}~\bibnamefont {P\'erez~Galv\'an}},
  \bibinfo {author} {\bibfnamefont {M.~L.}\ \bibnamefont {Pitt}}, \bibinfo
  {author} {\bibfnamefont {B.}~\bibnamefont {Plaster}}, \bibinfo {author}
  {\bibfnamefont {J.~C.}\ \bibnamefont {Ramsey}}, \bibinfo {author}
  {\bibfnamefont {R.}~\bibnamefont {Rios}}, \bibinfo {author} {\bibfnamefont
  {R.}~\bibnamefont {Russell}}, \bibinfo {author} {\bibfnamefont
  {A.}~\bibnamefont {Saunders}}, \bibinfo {author} {\bibfnamefont {S.~J.}\
  \bibnamefont {Seestrom}}, \bibinfo {author} {\bibfnamefont {W.~E.}\
  \bibnamefont {Sondheim}}, \bibinfo {author} {\bibfnamefont {E.}~\bibnamefont
  {Tatar}}, \bibinfo {author} {\bibfnamefont {R.~B.}\ \bibnamefont {Vogelaar}},
  \bibinfo {author} {\bibfnamefont {B.}~\bibnamefont {VornDick}}, \bibinfo
  {author} {\bibfnamefont {C.}~\bibnamefont {Wrede}}, \bibinfo {author}
  {\bibfnamefont {H.}~\bibnamefont {Yan}}, \bibinfo {author} {\bibfnamefont
  {A.~R.}\ \bibnamefont {Young}}, \ and\ \bibinfo {author} {\bibfnamefont
  {B.}~\bibnamefont {Zeck}} (\bibinfo {collaboration} {UCNA Collaboration}),\
  }\href {\doibase 10.1103/PhysRevLett.105.181803} {\bibfield  {journal}
  {\bibinfo  {journal} {Phys. Rev. Lett.}\ }\textbf {\bibinfo {volume} {105}},\
  \bibinfo {pages} {181803} (\bibinfo {year} {2010})}\BibitemShut {NoStop}%
\bibitem [{\citenamefont {Beringer}\ \emph {et~al.}(2012)\citenamefont
  {Beringer}, \citenamefont {Arguin}, \citenamefont {Barnett}, \citenamefont
  {Copic}, \citenamefont {Dahl}, \citenamefont {Groom}, \citenamefont {Lin},
  \citenamefont {Lys}, \citenamefont {Murayama}, \citenamefont {Wohl},
  \citenamefont {Yao}, \citenamefont {Zyla}, \citenamefont {Amsler},
  \citenamefont {Antonelli}, \citenamefont {Asner}, \citenamefont {Baer},
  \citenamefont {Band}, \citenamefont {Basaglia}, \citenamefont {Bauer},
  \citenamefont {Beatty}, \citenamefont {Belousov}, \citenamefont {Bergren},
  \citenamefont {Bernardi}, \citenamefont {Bertl}, \citenamefont {Bethke},
  \citenamefont {Bichsel}, \citenamefont {Biebel}, \citenamefont {Blucher},
  \citenamefont {Blusk}, \citenamefont {Brooijmans}, \citenamefont
  {Buchmueller}, \citenamefont {Cahn}, \citenamefont {Carena}, \citenamefont
  {Ceccucci}, \citenamefont {Chakraborty}, \citenamefont {Chen}, \citenamefont
  {Chivukula}, \citenamefont {Cowan}, \citenamefont {D'Ambrosio}, \citenamefont
  {Damour}, \citenamefont {de~Florian}, \citenamefont {de~Gouv\^ea},
  \citenamefont {DeGrand}, \citenamefont {de~Jong}, \citenamefont {Dissertori},
  \citenamefont {Dobrescu}, \citenamefont {Doser}, \citenamefont {Drees},
  \citenamefont {Edwards}, \citenamefont {Eidelman}, \citenamefont {Erler},
  \citenamefont {Ezhela}, \citenamefont {Fetscher}, \citenamefont {Fields},
  \citenamefont {Foster}, \citenamefont {Gaisser}, \citenamefont {Garren},
  \citenamefont {Gerber}, \citenamefont {Gerbier}, \citenamefont {Gherghetta},
  \citenamefont {Golwala}, \citenamefont {Goodman}, \citenamefont {Grab},
  \citenamefont {Gritsan}, \citenamefont {Grivaz}, \citenamefont {Gr\"unewald},
  \citenamefont {Gurtu}, \citenamefont {Gutsche}, \citenamefont {Haber},
  \citenamefont {Hagiwara}, \citenamefont {Hagmann}, \citenamefont {Hanhart},
  \citenamefont {Hashimoto}, \citenamefont {Hayes}, \citenamefont {Heffner},
  \citenamefont {Heltsley}, \citenamefont {Hern\'andez-Rey}, \citenamefont
  {Hikasa}, \citenamefont {H\"ocker}, \citenamefont {Holder}, \citenamefont
  {Holtkamp}, \citenamefont {Huston}, \citenamefont {Jackson}, \citenamefont
  {Johnson}, \citenamefont {Junk}, \citenamefont {Karlen}, \citenamefont
  {Kirkby}, \citenamefont {Klein}, \citenamefont {Klempt}, \citenamefont
  {Kowalewski}, \citenamefont {Krauss}, \citenamefont {Kreps}, \citenamefont
  {Krusche}, \citenamefont {Kuyanov}, \citenamefont {Kwon}, \citenamefont
  {Lahav}, \citenamefont {Laiho}, \citenamefont {Langacker}, \citenamefont
  {Liddle}, \citenamefont {Ligeti}, \citenamefont {Liss}, \citenamefont
  {Littenberg}, \citenamefont {Lugovsky}, \citenamefont {Lugovsky},
  \citenamefont {Mannel}, \citenamefont {Manohar}, \citenamefont {Marciano},
  \citenamefont {Martin}, \citenamefont {Masoni}, \citenamefont {Matthews},
  \citenamefont {Milstead}, \citenamefont {Miquel}, \citenamefont {M\"onig},
  \citenamefont {Moortgat}, \citenamefont {Nakamura}, \citenamefont {Narain},
  \citenamefont {Nason}, \citenamefont {Navas}, \citenamefont {Neubert},
  \citenamefont {Nevski}, \citenamefont {Nir}, \citenamefont {Olive},
  \citenamefont {Pape}, \citenamefont {Parsons}, \citenamefont {Patrignani},
  \citenamefont {Peacock}, \citenamefont {Petcov}, \citenamefont {Piepke},
  \citenamefont {Pomarol}, \citenamefont {Punzi}, \citenamefont {Quadt},
  \citenamefont {Raby}, \citenamefont {Raffelt}, \citenamefont {Ratcliff},
  \citenamefont {Richardson}, \citenamefont {Roesler}, \citenamefont {Rolli},
  \citenamefont {Romaniouk}, \citenamefont {Rosenberg}, \citenamefont {Rosner},
  \citenamefont {Sachrajda}, \citenamefont {Sakai}, \citenamefont {Salam},
  \citenamefont {Sarkar}, \citenamefont {Sauli}, \citenamefont {Schneider},
  \citenamefont {Scholberg}, \citenamefont {Scott}, \citenamefont {Seligman},
  \citenamefont {Shaevitz}, \citenamefont {Sharpe}, \citenamefont {Silari},
  \citenamefont {Sj\"ostrand}, \citenamefont {Skands}, \citenamefont {Smith},
  \citenamefont {Smoot}, \citenamefont {Spanier}, \citenamefont {Spieler},
  \citenamefont {Stahl}, \citenamefont {Stanev}, \citenamefont {Stone},
  \citenamefont {Sumiyoshi}, \citenamefont {Syphers}, \citenamefont
  {Takahashi}, \citenamefont {Tanabashi}, \citenamefont {Terning},
  \citenamefont {Titov}, \citenamefont {Tkachenko}, \citenamefont
  {T\"ornqvist}, \citenamefont {Tovey}, \citenamefont {Valencia}, \citenamefont
  {van Bibber}, \citenamefont {Venanzoni}, \citenamefont {Vincter},
  \citenamefont {Vogel}, \citenamefont {Vogt}, \citenamefont {Walkowiak},
  \citenamefont {Walter}, \citenamefont {Ward}, \citenamefont {Watari},
  \citenamefont {Weiglein}, \citenamefont {Weinberg}, \citenamefont {Wiencke},
  \citenamefont {Wolfenstein}, \citenamefont {Womersley}, \citenamefont
  {Woody}, \citenamefont {Workman}, \citenamefont {Yamamoto}, \citenamefont
  {Zeller}, \citenamefont {Zenin}, \citenamefont {Zhang}, \citenamefont {Zhu},
  \citenamefont {Harper}, \citenamefont {Lugovsky},\ and\ \citenamefont
  {Schaffner}}]{pdg:12}%
  \BibitemOpen
  \bibfield  {author} {\bibinfo {author} {\bibfnamefont {J.}~\bibnamefont
  {Beringer}}, \bibinfo {author} {\bibfnamefont {J.~F.}\ \bibnamefont
  {Arguin}}, \bibinfo {author} {\bibfnamefont {R.~M.}\ \bibnamefont {Barnett}},
  \bibinfo {author} {\bibfnamefont {K.}~\bibnamefont {Copic}}, \bibinfo
  {author} {\bibfnamefont {O.}~\bibnamefont {Dahl}}, \bibinfo {author}
  {\bibfnamefont {D.~E.}\ \bibnamefont {Groom}}, \bibinfo {author}
  {\bibfnamefont {C.~J.}\ \bibnamefont {Lin}}, \bibinfo {author} {\bibfnamefont
  {J.}~\bibnamefont {Lys}}, \bibinfo {author} {\bibfnamefont {H.}~\bibnamefont
  {Murayama}}, \bibinfo {author} {\bibfnamefont {C.~G.}\ \bibnamefont {Wohl}},
  \bibinfo {author} {\bibfnamefont {W.~M.}\ \bibnamefont {Yao}}, \bibinfo
  {author} {\bibfnamefont {P.~A.}\ \bibnamefont {Zyla}}, \bibinfo {author}
  {\bibfnamefont {C.}~\bibnamefont {Amsler}}, \bibinfo {author} {\bibfnamefont
  {M.}~\bibnamefont {Antonelli}}, \bibinfo {author} {\bibfnamefont {D.~M.}\
  \bibnamefont {Asner}}, \bibinfo {author} {\bibfnamefont {H.}~\bibnamefont
  {Baer}}, \bibinfo {author} {\bibfnamefont {H.~R.}\ \bibnamefont {Band}},
  \bibinfo {author} {\bibfnamefont {T.}~\bibnamefont {Basaglia}}, \bibinfo
  {author} {\bibfnamefont {C.~W.}\ \bibnamefont {Bauer}}, \bibinfo {author}
  {\bibfnamefont {J.~J.}\ \bibnamefont {Beatty}}, \bibinfo {author}
  {\bibfnamefont {V.~I.}\ \bibnamefont {Belousov}}, \bibinfo {author}
  {\bibfnamefont {E.}~\bibnamefont {Bergren}}, \bibinfo {author} {\bibfnamefont
  {G.}~\bibnamefont {Bernardi}}, \bibinfo {author} {\bibfnamefont
  {W.}~\bibnamefont {Bertl}}, \bibinfo {author} {\bibfnamefont
  {S.}~\bibnamefont {Bethke}}, \bibinfo {author} {\bibfnamefont
  {H.}~\bibnamefont {Bichsel}}, \bibinfo {author} {\bibfnamefont
  {O.}~\bibnamefont {Biebel}}, \bibinfo {author} {\bibfnamefont
  {E.}~\bibnamefont {Blucher}}, \bibinfo {author} {\bibfnamefont
  {S.}~\bibnamefont {Blusk}}, \bibinfo {author} {\bibfnamefont
  {G.}~\bibnamefont {Brooijmans}}, \bibinfo {author} {\bibfnamefont
  {O.}~\bibnamefont {Buchmueller}}, \bibinfo {author} {\bibfnamefont {R.~N.}\
  \bibnamefont {Cahn}}, \bibinfo {author} {\bibfnamefont {M.}~\bibnamefont
  {Carena}}, \bibinfo {author} {\bibfnamefont {A.}~\bibnamefont {Ceccucci}},
  \bibinfo {author} {\bibfnamefont {D.}~\bibnamefont {Chakraborty}}, \bibinfo
  {author} {\bibfnamefont {M.~C.}\ \bibnamefont {Chen}}, \bibinfo {author}
  {\bibfnamefont {R.~S.}\ \bibnamefont {Chivukula}}, \bibinfo {author}
  {\bibfnamefont {G.}~\bibnamefont {Cowan}}, \bibinfo {author} {\bibfnamefont
  {G.}~\bibnamefont {D'Ambrosio}}, \bibinfo {author} {\bibfnamefont
  {T.}~\bibnamefont {Damour}}, \bibinfo {author} {\bibfnamefont
  {D.}~\bibnamefont {de~Florian}}, \bibinfo {author} {\bibfnamefont
  {A.}~\bibnamefont {de~Gouv\^ea}}, \bibinfo {author} {\bibfnamefont
  {T.}~\bibnamefont {DeGrand}}, \bibinfo {author} {\bibfnamefont
  {P.}~\bibnamefont {de~Jong}}, \bibinfo {author} {\bibfnamefont
  {G.}~\bibnamefont {Dissertori}}, \bibinfo {author} {\bibfnamefont
  {B.}~\bibnamefont {Dobrescu}}, \bibinfo {author} {\bibfnamefont
  {M.}~\bibnamefont {Doser}}, \bibinfo {author} {\bibfnamefont
  {M.}~\bibnamefont {Drees}}, \bibinfo {author} {\bibfnamefont {D.~A.}\
  \bibnamefont {Edwards}}, \bibinfo {author} {\bibfnamefont {S.}~\bibnamefont
  {Eidelman}}, \bibinfo {author} {\bibfnamefont {J.}~\bibnamefont {Erler}},
  \bibinfo {author} {\bibfnamefont {V.~V.}\ \bibnamefont {Ezhela}}, \bibinfo
  {author} {\bibfnamefont {W.}~\bibnamefont {Fetscher}}, \bibinfo {author}
  {\bibfnamefont {B.~D.}\ \bibnamefont {Fields}}, \bibinfo {author}
  {\bibfnamefont {B.}~\bibnamefont {Foster}}, \bibinfo {author} {\bibfnamefont
  {T.~K.}\ \bibnamefont {Gaisser}}, \bibinfo {author} {\bibfnamefont
  {L.}~\bibnamefont {Garren}}, \bibinfo {author} {\bibfnamefont {H.~J.}\
  \bibnamefont {Gerber}}, \bibinfo {author} {\bibfnamefont {G.}~\bibnamefont
  {Gerbier}}, \bibinfo {author} {\bibfnamefont {T.}~\bibnamefont {Gherghetta}},
  \bibinfo {author} {\bibfnamefont {S.}~\bibnamefont {Golwala}}, \bibinfo
  {author} {\bibfnamefont {M.}~\bibnamefont {Goodman}}, \bibinfo {author}
  {\bibfnamefont {C.}~\bibnamefont {Grab}}, \bibinfo {author} {\bibfnamefont
  {A.~V.}\ \bibnamefont {Gritsan}}, \bibinfo {author} {\bibfnamefont {J.~F.}\
  \bibnamefont {Grivaz}}, \bibinfo {author} {\bibfnamefont {M.}~\bibnamefont
  {Gr\"unewald}}, \bibinfo {author} {\bibfnamefont {A.}~\bibnamefont {Gurtu}},
  \bibinfo {author} {\bibfnamefont {T.}~\bibnamefont {Gutsche}}, \bibinfo
  {author} {\bibfnamefont {H.~E.}\ \bibnamefont {Haber}}, \bibinfo {author}
  {\bibfnamefont {K.}~\bibnamefont {Hagiwara}}, \bibinfo {author}
  {\bibfnamefont {C.}~\bibnamefont {Hagmann}}, \bibinfo {author} {\bibfnamefont
  {C.}~\bibnamefont {Hanhart}}, \bibinfo {author} {\bibfnamefont
  {S.}~\bibnamefont {Hashimoto}}, \bibinfo {author} {\bibfnamefont {K.~G.}\
  \bibnamefont {Hayes}}, \bibinfo {author} {\bibfnamefont {M.}~\bibnamefont
  {Heffner}}, \bibinfo {author} {\bibfnamefont {B.}~\bibnamefont {Heltsley}},
  \bibinfo {author} {\bibfnamefont {J.~J.}\ \bibnamefont {Hern\'andez-Rey}},
  \bibinfo {author} {\bibfnamefont {K.}~\bibnamefont {Hikasa}}, \bibinfo
  {author} {\bibfnamefont {A.}~\bibnamefont {H\"ocker}}, \bibinfo {author}
  {\bibfnamefont {J.}~\bibnamefont {Holder}}, \bibinfo {author} {\bibfnamefont
  {A.}~\bibnamefont {Holtkamp}}, \bibinfo {author} {\bibfnamefont
  {J.}~\bibnamefont {Huston}}, \bibinfo {author} {\bibfnamefont {J.~D.}\
  \bibnamefont {Jackson}}, \bibinfo {author} {\bibfnamefont {K.~F.}\
  \bibnamefont {Johnson}}, \bibinfo {author} {\bibfnamefont {T.}~\bibnamefont
  {Junk}}, \bibinfo {author} {\bibfnamefont {D.}~\bibnamefont {Karlen}},
  \bibinfo {author} {\bibfnamefont {D.}~\bibnamefont {Kirkby}}, \bibinfo
  {author} {\bibfnamefont {S.~R.}\ \bibnamefont {Klein}}, \bibinfo {author}
  {\bibfnamefont {E.}~\bibnamefont {Klempt}}, \bibinfo {author} {\bibfnamefont
  {R.~V.}\ \bibnamefont {Kowalewski}}, \bibinfo {author} {\bibfnamefont
  {F.}~\bibnamefont {Krauss}}, \bibinfo {author} {\bibfnamefont
  {M.}~\bibnamefont {Kreps}}, \bibinfo {author} {\bibfnamefont
  {B.}~\bibnamefont {Krusche}}, \bibinfo {author} {\bibfnamefont {Y.~V.}\
  \bibnamefont {Kuyanov}}, \bibinfo {author} {\bibfnamefont {Y.}~\bibnamefont
  {Kwon}}, \bibinfo {author} {\bibfnamefont {O.}~\bibnamefont {Lahav}},
  \bibinfo {author} {\bibfnamefont {J.}~\bibnamefont {Laiho}}, \bibinfo
  {author} {\bibfnamefont {P.}~\bibnamefont {Langacker}}, \bibinfo {author}
  {\bibfnamefont {A.}~\bibnamefont {Liddle}}, \bibinfo {author} {\bibfnamefont
  {Z.}~\bibnamefont {Ligeti}}, \bibinfo {author} {\bibfnamefont {T.~M.}\
  \bibnamefont {Liss}}, \bibinfo {author} {\bibfnamefont {L.}~\bibnamefont
  {Littenberg}}, \bibinfo {author} {\bibfnamefont {K.~S.}\ \bibnamefont
  {Lugovsky}}, \bibinfo {author} {\bibfnamefont {S.~B.}\ \bibnamefont
  {Lugovsky}}, \bibinfo {author} {\bibfnamefont {T.}~\bibnamefont {Mannel}},
  \bibinfo {author} {\bibfnamefont {A.~V.}\ \bibnamefont {Manohar}}, \bibinfo
  {author} {\bibfnamefont {W.~J.}\ \bibnamefont {Marciano}}, \bibinfo {author}
  {\bibfnamefont {A.~D.}\ \bibnamefont {Martin}}, \bibinfo {author}
  {\bibfnamefont {A.}~\bibnamefont {Masoni}}, \bibinfo {author} {\bibfnamefont
  {J.}~\bibnamefont {Matthews}}, \bibinfo {author} {\bibfnamefont
  {D.}~\bibnamefont {Milstead}}, \bibinfo {author} {\bibfnamefont
  {R.}~\bibnamefont {Miquel}}, \bibinfo {author} {\bibfnamefont
  {K.}~\bibnamefont {M\"onig}}, \bibinfo {author} {\bibfnamefont
  {F.}~\bibnamefont {Moortgat}}, \bibinfo {author} {\bibfnamefont
  {K.}~\bibnamefont {Nakamura}}, \bibinfo {author} {\bibfnamefont
  {M.}~\bibnamefont {Narain}}, \bibinfo {author} {\bibfnamefont
  {P.}~\bibnamefont {Nason}}, \bibinfo {author} {\bibfnamefont
  {S.}~\bibnamefont {Navas}}, \bibinfo {author} {\bibfnamefont
  {M.}~\bibnamefont {Neubert}}, \bibinfo {author} {\bibfnamefont
  {P.}~\bibnamefont {Nevski}}, \bibinfo {author} {\bibfnamefont
  {Y.}~\bibnamefont {Nir}}, \bibinfo {author} {\bibfnamefont {K.~A.}\
  \bibnamefont {Olive}}, \bibinfo {author} {\bibfnamefont {L.}~\bibnamefont
  {Pape}}, \bibinfo {author} {\bibfnamefont {J.}~\bibnamefont {Parsons}},
  \bibinfo {author} {\bibfnamefont {C.}~\bibnamefont {Patrignani}}, \bibinfo
  {author} {\bibfnamefont {J.~A.}\ \bibnamefont {Peacock}}, \bibinfo {author}
  {\bibfnamefont {S.~T.}\ \bibnamefont {Petcov}}, \bibinfo {author}
  {\bibfnamefont {A.}~\bibnamefont {Piepke}}, \bibinfo {author} {\bibfnamefont
  {A.}~\bibnamefont {Pomarol}}, \bibinfo {author} {\bibfnamefont
  {G.}~\bibnamefont {Punzi}}, \bibinfo {author} {\bibfnamefont
  {A.}~\bibnamefont {Quadt}}, \bibinfo {author} {\bibfnamefont
  {S.}~\bibnamefont {Raby}}, \bibinfo {author} {\bibfnamefont {G.}~\bibnamefont
  {Raffelt}}, \bibinfo {author} {\bibfnamefont {B.~N.}\ \bibnamefont
  {Ratcliff}}, \bibinfo {author} {\bibfnamefont {P.}~\bibnamefont
  {Richardson}}, \bibinfo {author} {\bibfnamefont {S.}~\bibnamefont {Roesler}},
  \bibinfo {author} {\bibfnamefont {S.}~\bibnamefont {Rolli}}, \bibinfo
  {author} {\bibfnamefont {A.}~\bibnamefont {Romaniouk}}, \bibinfo {author}
  {\bibfnamefont {L.~J.}\ \bibnamefont {Rosenberg}}, \bibinfo {author}
  {\bibfnamefont {J.~L.}\ \bibnamefont {Rosner}}, \bibinfo {author}
  {\bibfnamefont {C.~T.}\ \bibnamefont {Sachrajda}}, \bibinfo {author}
  {\bibfnamefont {Y.}~\bibnamefont {Sakai}}, \bibinfo {author} {\bibfnamefont
  {G.~P.}\ \bibnamefont {Salam}}, \bibinfo {author} {\bibfnamefont
  {S.}~\bibnamefont {Sarkar}}, \bibinfo {author} {\bibfnamefont
  {F.}~\bibnamefont {Sauli}}, \bibinfo {author} {\bibfnamefont
  {O.}~\bibnamefont {Schneider}}, \bibinfo {author} {\bibfnamefont
  {K.}~\bibnamefont {Scholberg}}, \bibinfo {author} {\bibfnamefont
  {D.}~\bibnamefont {Scott}}, \bibinfo {author} {\bibfnamefont {W.~G.}\
  \bibnamefont {Seligman}}, \bibinfo {author} {\bibfnamefont {M.~H.}\
  \bibnamefont {Shaevitz}}, \bibinfo {author} {\bibfnamefont {S.~R.}\
  \bibnamefont {Sharpe}}, \bibinfo {author} {\bibfnamefont {M.}~\bibnamefont
  {Silari}}, \bibinfo {author} {\bibfnamefont {T.}~\bibnamefont {Sj\"ostrand}},
  \bibinfo {author} {\bibfnamefont {P.}~\bibnamefont {Skands}}, \bibinfo
  {author} {\bibfnamefont {J.~G.}\ \bibnamefont {Smith}}, \bibinfo {author}
  {\bibfnamefont {G.~F.}\ \bibnamefont {Smoot}}, \bibinfo {author}
  {\bibfnamefont {S.}~\bibnamefont {Spanier}}, \bibinfo {author} {\bibfnamefont
  {H.}~\bibnamefont {Spieler}}, \bibinfo {author} {\bibfnamefont
  {A.}~\bibnamefont {Stahl}}, \bibinfo {author} {\bibfnamefont
  {T.}~\bibnamefont {Stanev}}, \bibinfo {author} {\bibfnamefont {S.~L.}\
  \bibnamefont {Stone}}, \bibinfo {author} {\bibfnamefont {T.}~\bibnamefont
  {Sumiyoshi}}, \bibinfo {author} {\bibfnamefont {M.~J.}\ \bibnamefont
  {Syphers}}, \bibinfo {author} {\bibfnamefont {F.}~\bibnamefont {Takahashi}},
  \bibinfo {author} {\bibfnamefont {M.}~\bibnamefont {Tanabashi}}, \bibinfo
  {author} {\bibfnamefont {J.}~\bibnamefont {Terning}}, \bibinfo {author}
  {\bibfnamefont {M.}~\bibnamefont {Titov}}, \bibinfo {author} {\bibfnamefont
  {N.~P.}\ \bibnamefont {Tkachenko}}, \bibinfo {author} {\bibfnamefont {N.~A.}\
  \bibnamefont {T\"ornqvist}}, \bibinfo {author} {\bibfnamefont
  {D.}~\bibnamefont {Tovey}}, \bibinfo {author} {\bibfnamefont
  {G.}~\bibnamefont {Valencia}}, \bibinfo {author} {\bibfnamefont
  {K.}~\bibnamefont {van Bibber}}, \bibinfo {author} {\bibfnamefont
  {G.}~\bibnamefont {Venanzoni}}, \bibinfo {author} {\bibfnamefont {M.~G.}\
  \bibnamefont {Vincter}}, \bibinfo {author} {\bibfnamefont {P.}~\bibnamefont
  {Vogel}}, \bibinfo {author} {\bibfnamefont {A.}~\bibnamefont {Vogt}},
  \bibinfo {author} {\bibfnamefont {W.}~\bibnamefont {Walkowiak}}, \bibinfo
  {author} {\bibfnamefont {C.~W.}\ \bibnamefont {Walter}}, \bibinfo {author}
  {\bibfnamefont {D.~R.}\ \bibnamefont {Ward}}, \bibinfo {author}
  {\bibfnamefont {T.}~\bibnamefont {Watari}}, \bibinfo {author} {\bibfnamefont
  {G.}~\bibnamefont {Weiglein}}, \bibinfo {author} {\bibfnamefont {E.~J.}\
  \bibnamefont {Weinberg}}, \bibinfo {author} {\bibfnamefont {L.~R.}\
  \bibnamefont {Wiencke}}, \bibinfo {author} {\bibfnamefont {L.}~\bibnamefont
  {Wolfenstein}}, \bibinfo {author} {\bibfnamefont {J.}~\bibnamefont
  {Womersley}}, \bibinfo {author} {\bibfnamefont {C.~L.}\ \bibnamefont
  {Woody}}, \bibinfo {author} {\bibfnamefont {R.~L.}\ \bibnamefont {Workman}},
  \bibinfo {author} {\bibfnamefont {A.}~\bibnamefont {Yamamoto}}, \bibinfo
  {author} {\bibfnamefont {G.~P.}\ \bibnamefont {Zeller}}, \bibinfo {author}
  {\bibfnamefont {O.~V.}\ \bibnamefont {Zenin}}, \bibinfo {author}
  {\bibfnamefont {J.}~\bibnamefont {Zhang}}, \bibinfo {author} {\bibfnamefont
  {R.~Y.}\ \bibnamefont {Zhu}}, \bibinfo {author} {\bibfnamefont
  {G.}~\bibnamefont {Harper}}, \bibinfo {author} {\bibfnamefont {V.~S.}\
  \bibnamefont {Lugovsky}}, \ and\ \bibinfo {author} {\bibfnamefont
  {P.}~\bibnamefont {Schaffner}} (\bibinfo {collaboration} {Particle Data
  Group}),\ }\href {\doibase 10.1103/PhysRevD.86.010001} {\bibfield  {journal}
  {\bibinfo  {journal} {Phys. Rev. D}\ }\textbf {\bibinfo {volume} {86}},\
  \bibinfo {pages} {010001} (\bibinfo {year} {2012})}\BibitemShut {NoStop}%
\bibitem [{\citenamefont {Mendenhall}\ \emph {et~al.}(2013)\citenamefont
  {Mendenhall}, \citenamefont {Pattie}, \citenamefont {Bagdasarova},
  \citenamefont {Berguno}, \citenamefont {Broussard}, \citenamefont {Carr},
  \citenamefont {Currie}, \citenamefont {Ding}, \citenamefont {Filippone},
  \citenamefont {Garc\'ia}, \citenamefont {Geltenbort}, \citenamefont
  {Hickerson}, \citenamefont {Hoagland}, \citenamefont {Holley}, \citenamefont
  {Hong}, \citenamefont {Ito}, \citenamefont {Knecht}, \citenamefont {Liu},
  \citenamefont {Liu}, \citenamefont {Makela}, \citenamefont {Mammei},
  \citenamefont {Martin}, \citenamefont {Melconian}, \citenamefont {Moore},
  \citenamefont {Morris}, \citenamefont {P\'erez~Galv\'an}, \citenamefont
  {Picker}, \citenamefont {Pitt}, \citenamefont {Plaster}, \citenamefont
  {Ramsey}, \citenamefont {Rios}, \citenamefont {Saunders}, \citenamefont
  {Seestrom}, \citenamefont {Sharapov}, \citenamefont {Sondheim}, \citenamefont
  {Tatar}, \citenamefont {Vogelaar}, \citenamefont {VornDick}, \citenamefont
  {Wrede}, \citenamefont {Young},\ and\ \citenamefont {Zeck}}]{Mendenhall2013}%
  \BibitemOpen
  \bibfield  {author} {\bibinfo {author} {\bibfnamefont {M.~P.}\ \bibnamefont
  {Mendenhall}}, \bibinfo {author} {\bibfnamefont {R.~W.}\ \bibnamefont
  {Pattie}}, \bibinfo {author} {\bibfnamefont {Y.}~\bibnamefont {Bagdasarova}},
  \bibinfo {author} {\bibfnamefont {D.~B.}\ \bibnamefont {Berguno}}, \bibinfo
  {author} {\bibfnamefont {L.~J.}\ \bibnamefont {Broussard}}, \bibinfo {author}
  {\bibfnamefont {R.}~\bibnamefont {Carr}}, \bibinfo {author} {\bibfnamefont
  {S.}~\bibnamefont {Currie}}, \bibinfo {author} {\bibfnamefont
  {X.}~\bibnamefont {Ding}}, \bibinfo {author} {\bibfnamefont {B.~W.}\
  \bibnamefont {Filippone}}, \bibinfo {author} {\bibfnamefont {A.}~\bibnamefont
  {Garc\'ia}}, \bibinfo {author} {\bibfnamefont {P.}~\bibnamefont
  {Geltenbort}}, \bibinfo {author} {\bibfnamefont {K.~P.}\ \bibnamefont
  {Hickerson}}, \bibinfo {author} {\bibfnamefont {J.}~\bibnamefont {Hoagland}},
  \bibinfo {author} {\bibfnamefont {A.~T.}\ \bibnamefont {Holley}}, \bibinfo
  {author} {\bibfnamefont {R.}~\bibnamefont {Hong}}, \bibinfo {author}
  {\bibfnamefont {T.~M.}\ \bibnamefont {Ito}}, \bibinfo {author} {\bibfnamefont
  {A.}~\bibnamefont {Knecht}}, \bibinfo {author} {\bibfnamefont {C.-Y.}\
  \bibnamefont {Liu}}, \bibinfo {author} {\bibfnamefont {J.~L.}\ \bibnamefont
  {Liu}}, \bibinfo {author} {\bibfnamefont {M.}~\bibnamefont {Makela}},
  \bibinfo {author} {\bibfnamefont {R.~R.}\ \bibnamefont {Mammei}}, \bibinfo
  {author} {\bibfnamefont {J.~W.}\ \bibnamefont {Martin}}, \bibinfo {author}
  {\bibfnamefont {D.}~\bibnamefont {Melconian}}, \bibinfo {author}
  {\bibfnamefont {S.~D.}\ \bibnamefont {Moore}}, \bibinfo {author}
  {\bibfnamefont {C.~L.}\ \bibnamefont {Morris}}, \bibinfo {author}
  {\bibfnamefont {A.}~\bibnamefont {P\'erez~Galv\'an}}, \bibinfo {author}
  {\bibfnamefont {R.}~\bibnamefont {Picker}}, \bibinfo {author} {\bibfnamefont
  {M.~L.}\ \bibnamefont {Pitt}}, \bibinfo {author} {\bibfnamefont
  {B.}~\bibnamefont {Plaster}}, \bibinfo {author} {\bibfnamefont {J.~C.}\
  \bibnamefont {Ramsey}}, \bibinfo {author} {\bibfnamefont {R.}~\bibnamefont
  {Rios}}, \bibinfo {author} {\bibfnamefont {A.}~\bibnamefont {Saunders}},
  \bibinfo {author} {\bibfnamefont {S.~J.}\ \bibnamefont {Seestrom}}, \bibinfo
  {author} {\bibfnamefont {E.~I.}\ \bibnamefont {Sharapov}}, \bibinfo {author}
  {\bibfnamefont {W.~E.}\ \bibnamefont {Sondheim}}, \bibinfo {author}
  {\bibfnamefont {E.}~\bibnamefont {Tatar}}, \bibinfo {author} {\bibfnamefont
  {R.~B.}\ \bibnamefont {Vogelaar}}, \bibinfo {author} {\bibfnamefont
  {B.}~\bibnamefont {VornDick}}, \bibinfo {author} {\bibfnamefont
  {C.}~\bibnamefont {Wrede}}, \bibinfo {author} {\bibfnamefont {A.~R.}\
  \bibnamefont {Young}}, \ and\ \bibinfo {author} {\bibfnamefont {B.~A.}\
  \bibnamefont {Zeck}} (\bibinfo {collaboration} {UCNA Collaboration}),\
  }\href@noop {} {\bibfield  {journal} {\bibinfo  {journal} {Phys. Rev. C}\
  }\textbf {\bibinfo {volume} {87}},\ \bibinfo {pages} {032501} (\bibinfo
  {year} {2013})}\BibitemShut {NoStop}%
\bibitem [{\citenamefont {Mund}\ \emph {et~al.}(2013)\citenamefont {Mund},
  \citenamefont {M\"arkisch}, \citenamefont {Deissenroth}, \citenamefont
  {Krempel}, \citenamefont {Schumann}, \citenamefont {Abele}, \citenamefont
  {Petoukhov},\ and\ \citenamefont {Soldner}}]{Mund2013}%
  \BibitemOpen
  \bibfield  {author} {\bibinfo {author} {\bibfnamefont {D.}~\bibnamefont
  {Mund}}, \bibinfo {author} {\bibfnamefont {B.}~\bibnamefont {M\"arkisch}},
  \bibinfo {author} {\bibfnamefont {M.}~\bibnamefont {Deissenroth}}, \bibinfo
  {author} {\bibfnamefont {J.}~\bibnamefont {Krempel}}, \bibinfo {author}
  {\bibfnamefont {M.}~\bibnamefont {Schumann}}, \bibinfo {author}
  {\bibfnamefont {H.}~\bibnamefont {Abele}}, \bibinfo {author} {\bibfnamefont
  {A.}~\bibnamefont {Petoukhov}}, \ and\ \bibinfo {author} {\bibfnamefont
  {T.}~\bibnamefont {Soldner}},\ }\href@noop {} {\bibfield  {journal} {\bibinfo
   {journal} {Phys. Rev. Lett.}\ }\textbf {\bibinfo {volume} {110}},\ \bibinfo
  {pages} {172502} (\bibinfo {year} {2013})}\BibitemShut {NoStop}%
\bibitem [{\citenamefont {Wilkinson}(1982)}]{Wilkinson1982}%
  \BibitemOpen
  \bibfield  {author} {\bibinfo {author} {\bibfnamefont {D.}~\bibnamefont
  {Wilkinson}},\ }\href {\doibase 10.1016/0375-9474(82)90051-3} {\bibfield
  {journal} {\bibinfo  {journal} {Nuclear Physics A}\ }\textbf {\bibinfo
  {volume} {377}},\ \bibinfo {pages} {474 } (\bibinfo {year}
  {1982})}\BibitemShut {NoStop}%
\bibitem [{\citenamefont {Gl\"uck}\ and\ \citenamefont
  {T\'oth}(1992)}]{Gluck1992}%
  \BibitemOpen
  \bibfield  {author} {\bibinfo {author} {\bibfnamefont {F.}~\bibnamefont
  {Gl\"uck}}\ and\ \bibinfo {author} {\bibfnamefont {K.}~\bibnamefont
  {T\'oth}},\ }\href {\doibase 10.1103/PhysRevD.46.2090} {\bibfield  {journal}
  {\bibinfo  {journal} {Phys. Rev. D}\ }\textbf {\bibinfo {volume} {46}},\
  \bibinfo {pages} {2090} (\bibinfo {year} {1992})}\BibitemShut {NoStop}%
\bibitem [{\citenamefont {Marciano}\ and\ \citenamefont
  {Sirlin}(2006)}]{Marciano2006}%
  \BibitemOpen
  \bibfield  {author} {\bibinfo {author} {\bibfnamefont {W.~J.}\ \bibnamefont
  {Marciano}}\ and\ \bibinfo {author} {\bibfnamefont {A.}~\bibnamefont
  {Sirlin}},\ }\href@noop {} {\bibfield  {journal} {\bibinfo  {journal} {Phys.
  Rev. Lett.}\ }\textbf {\bibinfo {volume} {96}},\ \bibinfo {pages} {032002}
  (\bibinfo {year} {2006})}\BibitemShut {NoStop}%
\bibitem [{\citenamefont {Wietfeldt}\ and\ \citenamefont
  {Greene}(2011)}]{Wietfeldt:2011}%
  \BibitemOpen
  \bibfield  {author} {\bibinfo {author} {\bibfnamefont {F.~E.}\ \bibnamefont
  {Wietfeldt}}\ and\ \bibinfo {author} {\bibfnamefont {G.~L.}\ \bibnamefont
  {Greene}},\ }\href@noop {} {\bibfield  {journal} {\bibinfo  {journal} {Rev.
  Mod. Phys.}\ }\textbf {\bibinfo {volume} {83}},\ \bibinfo {pages} {1173}
  (\bibinfo {year} {2011})}\BibitemShut {NoStop}%
\bibitem [{\citenamefont {Wauters}\ \emph {et~al.}(2010)\citenamefont
  {Wauters}, \citenamefont {Kraev}, \citenamefont {Z\'akouck\'y}, \citenamefont
  {Beck}, \citenamefont {Breitenfeldt}, \citenamefont {De~Leebeeck},
  \citenamefont {Golovko}, \citenamefont {Kozlov}, \citenamefont {Phalet},
  \citenamefont {Roccia}, \citenamefont {Soti}, \citenamefont {Tandecki},
  \citenamefont {Towner}, \citenamefont {Traykov}, \citenamefont {Van~Gorp},\
  and\ \citenamefont {Severijns}}]{Wauters:2010}%
  \BibitemOpen
  \bibfield  {author} {\bibinfo {author} {\bibfnamefont {F.}~\bibnamefont
  {Wauters}}, \bibinfo {author} {\bibfnamefont {I.}~\bibnamefont {Kraev}},
  \bibinfo {author} {\bibfnamefont {D.}~\bibnamefont {Z\'akouck\'y}}, \bibinfo
  {author} {\bibfnamefont {M.}~\bibnamefont {Beck}}, \bibinfo {author}
  {\bibfnamefont {M.}~\bibnamefont {Breitenfeldt}}, \bibinfo {author}
  {\bibfnamefont {V.}~\bibnamefont {De~Leebeeck}}, \bibinfo {author}
  {\bibfnamefont {V.~V.}\ \bibnamefont {Golovko}}, \bibinfo {author}
  {\bibfnamefont {V.~Y.}\ \bibnamefont {Kozlov}}, \bibinfo {author}
  {\bibfnamefont {T.}~\bibnamefont {Phalet}}, \bibinfo {author} {\bibfnamefont
  {S.}~\bibnamefont {Roccia}}, \bibinfo {author} {\bibfnamefont
  {G.}~\bibnamefont {Soti}}, \bibinfo {author} {\bibfnamefont {M.}~\bibnamefont
  {Tandecki}}, \bibinfo {author} {\bibfnamefont {I.~S.}\ \bibnamefont
  {Towner}}, \bibinfo {author} {\bibfnamefont {E.}~\bibnamefont {Traykov}},
  \bibinfo {author} {\bibfnamefont {S.}~\bibnamefont {Van~Gorp}}, \ and\
  \bibinfo {author} {\bibfnamefont {N.}~\bibnamefont {Severijns}},\ }\href
  {\doibase 10.1103/PhysRevC.82.055502} {\bibfield  {journal} {\bibinfo
  {journal} {Phys. Rev. C}\ }\textbf {\bibinfo {volume} {82}},\ \bibinfo
  {pages} {055502} (\bibinfo {year} {2010})}\BibitemShut {NoStop}%
\bibitem [{\citenamefont {Holstein}(1974)}]{Holstein1974}%
  \BibitemOpen
  \bibfield  {author} {\bibinfo {author} {\bibfnamefont {B.}~\bibnamefont
  {Holstein}},\ }\href@noop {} {\bibfield  {journal} {\bibinfo  {journal}
  {Reviews of Modern Physics}\ }\textbf {\bibinfo {volume} {46}},\ \bibinfo
  {pages} {789 } (\bibinfo {year} {1974})}\BibitemShut {NoStop}%
\bibitem [{\citenamefont {Adelberger}\ \emph {et~al.}(1999)\citenamefont
  {Adelberger}, \citenamefont {Ortiz}, \citenamefont {Garcia}, \citenamefont
  {Swanson}, \citenamefont {Beck}, \citenamefont {Tengblad}, \citenamefont
  {Borge}, \citenamefont {Martel}, \citenamefont {Bichsel},\ and\ \citenamefont
  {the ISOLDE~collaboration}}]{Adelberger1999}%
  \BibitemOpen
  \bibfield  {author} {\bibinfo {author} {\bibfnamefont {E.~G.}\ \bibnamefont
  {Adelberger}}, \bibinfo {author} {\bibfnamefont {C.}~\bibnamefont {Ortiz}},
  \bibinfo {author} {\bibfnamefont {A.}~\bibnamefont {Garcia}}, \bibinfo
  {author} {\bibfnamefont {H.~E.}\ \bibnamefont {Swanson}}, \bibinfo {author}
  {\bibfnamefont {M.}~\bibnamefont {Beck}}, \bibinfo {author} {\bibfnamefont
  {O.}~\bibnamefont {Tengblad}}, \bibinfo {author} {\bibfnamefont {M.~J.~G.}\
  \bibnamefont {Borge}}, \bibinfo {author} {\bibfnamefont {I.}~\bibnamefont
  {Martel}}, \bibinfo {author} {\bibfnamefont {H.}~\bibnamefont {Bichsel}}, \
  and\ \bibinfo {author} {\bibnamefont {the ISOLDE~collaboration}},\
  }\href@noop {} {\bibfield  {journal} {\bibinfo  {journal} {Physical Review
  Letters}\ }\textbf {\bibinfo {volume} {83}},\ \bibinfo {pages} {1299 }
  (\bibinfo {year} {1999})}\BibitemShut {NoStop}%
\bibitem [{\citenamefont {Johnson}\ \emph {et~al.}(1963)\citenamefont
  {Johnson}, \citenamefont {Pleasonton},\ and\ \citenamefont
  {Carlson}}]{Johnson1963}%
  \BibitemOpen
  \bibfield  {author} {\bibinfo {author} {\bibfnamefont {C.}~\bibnamefont
  {Johnson}}, \bibinfo {author} {\bibfnamefont {F.}~\bibnamefont {Pleasonton}},
  \ and\ \bibinfo {author} {\bibfnamefont {T.}~\bibnamefont {Carlson}},\
  }\href@noop {} {\bibfield  {journal} {\bibinfo  {journal} {Physical Review}\
  }\textbf {\bibinfo {volume} {132}},\ \bibinfo {pages} {1149} (\bibinfo {year}
  {1963})}\BibitemShut {NoStop}%
\bibitem [{\citenamefont {Gl\"uck}(1998)}]{Gluck:1998}%
  \BibitemOpen
  \bibfield  {author} {\bibinfo {author} {\bibfnamefont {F.}~\bibnamefont
  {Gl\"uck}},\ }\href {\doibase 10.1016/S0375-9474(97)00643-X", url
  ={http://www.sciencedirect.com/science/article/pii/S037594749700643X},}
  {\bibfield  {journal} {\bibinfo  {journal} {Nuclear Physics A}\ }\textbf
  {\bibinfo {volume} {628}},\ \bibinfo {pages} {493 } (\bibinfo {year}
  {1998})}\BibitemShut {NoStop}%
\bibitem [{\citenamefont {Gorelov}\ \emph {et~al.}(2005)\citenamefont
  {Gorelov}, \citenamefont {Melconian}, \citenamefont {Alford}, \citenamefont
  {Ashery}, \citenamefont {Ball}, \citenamefont {Behr}, \citenamefont
  {Bricault}, \citenamefont {D'Auria}, \citenamefont {Deutch}, \citenamefont
  {Dilling}, \citenamefont {Dombsky}, \citenamefont {Dub\'{e}}, \citenamefont
  {Fingler}, \citenamefont {Giesen}, \citenamefont {Gl\"uck}, \citenamefont
  {Gu}, \citenamefont {H\"{a}usser}, \citenamefont {Jackson}, \citenamefont
  {Jennings}, \citenamefont {Pearson}, \citenamefont {Stocki}, \citenamefont
  {Swanson},\ and\ \citenamefont {Trinczek}}]{Gorelov2005}%
  \BibitemOpen
  \bibfield  {author} {\bibinfo {author} {\bibfnamefont {A.}~\bibnamefont
  {Gorelov}}, \bibinfo {author} {\bibfnamefont {D.}~\bibnamefont {Melconian}},
  \bibinfo {author} {\bibfnamefont {W.~P.}\ \bibnamefont {Alford}}, \bibinfo
  {author} {\bibfnamefont {D.}~\bibnamefont {Ashery}}, \bibinfo {author}
  {\bibfnamefont {G.}~\bibnamefont {Ball}}, \bibinfo {author} {\bibfnamefont
  {J.}~\bibnamefont {Behr}}, \bibinfo {author} {\bibfnamefont {P.~G.}\
  \bibnamefont {Bricault}}, \bibinfo {author} {\bibfnamefont {J.}~\bibnamefont
  {D'Auria}}, \bibinfo {author} {\bibfnamefont {J.}~\bibnamefont {Deutch}},
  \bibinfo {author} {\bibfnamefont {J.}~\bibnamefont {Dilling}}, \bibinfo
  {author} {\bibfnamefont {M.}~\bibnamefont {Dombsky}}, \bibinfo {author}
  {\bibfnamefont {P.}~\bibnamefont {Dub\'{e}}}, \bibinfo {author}
  {\bibfnamefont {J.}~\bibnamefont {Fingler}}, \bibinfo {author} {\bibfnamefont
  {U.}~\bibnamefont {Giesen}}, \bibinfo {author} {\bibfnamefont
  {F.}~\bibnamefont {Gl\"uck}}, \bibinfo {author} {\bibfnamefont
  {S.}~\bibnamefont {Gu}}, \bibinfo {author} {\bibfnamefont {O.}~\bibnamefont
  {H\"{a}usser}}, \bibinfo {author} {\bibfnamefont {K.}~\bibnamefont
  {Jackson}}, \bibinfo {author} {\bibfnamefont {B.}~\bibnamefont {Jennings}},
  \bibinfo {author} {\bibfnamefont {M.~R.}\ \bibnamefont {Pearson}}, \bibinfo
  {author} {\bibfnamefont {T.~J.}\ \bibnamefont {Stocki}}, \bibinfo {author}
  {\bibfnamefont {T.}~\bibnamefont {Swanson}}, \ and\ \bibinfo {author}
  {\bibfnamefont {M.}~\bibnamefont {Trinczek}},\ }\href@noop {} {\bibfield
  {journal} {\bibinfo  {journal} {Physical Review Letters}\ }\textbf {\bibinfo
  {volume} {94}},\ \bibinfo {pages} {142501} (\bibinfo {year}
  {2005})}\BibitemShut {NoStop}%
\bibitem [{\citenamefont {Hardy}\ and\ \citenamefont
  {Towner}(2005)}]{Hardy2005}%
  \BibitemOpen
  \bibfield  {author} {\bibinfo {author} {\bibfnamefont {J.~C.}\ \bibnamefont
  {Hardy}}\ and\ \bibinfo {author} {\bibfnamefont {I.~S.}\ \bibnamefont
  {Towner}},\ }\href {\doibase 10.1103/PhysRevC.71.055501} {\bibfield
  {journal} {\bibinfo  {journal} {Phys. Rev. C}\ }\textbf {\bibinfo {volume}
  {71}},\ \bibinfo {pages} {055501} (\bibinfo {year} {2005})}\BibitemShut
  {NoStop}%
\bibitem [{\citenamefont {Pitcairn}\ \emph {et~al.}(2009)\citenamefont
  {Pitcairn}, \citenamefont {Roberge}, \citenamefont {Gorelov}, \citenamefont
  {Ashery}, \citenamefont {Aviv}, \citenamefont {Behr}, \citenamefont
  {Bricault}, \citenamefont {Dombsky}, \citenamefont {Holt}, \citenamefont
  {Jackson}, \citenamefont {Lee}, \citenamefont {Pearson}, \citenamefont
  {Gaudin}, \citenamefont {Dej}, \citenamefont {H\"{o}hr}, \citenamefont
  {Gwinner},\ and\ \citenamefont {Melconian}}]{Pitcairn2009}%
  \BibitemOpen
  \bibfield  {author} {\bibinfo {author} {\bibfnamefont {J.~R.~A.}\
  \bibnamefont {Pitcairn}}, \bibinfo {author} {\bibfnamefont {D.}~\bibnamefont
  {Roberge}}, \bibinfo {author} {\bibfnamefont {A.}~\bibnamefont {Gorelov}},
  \bibinfo {author} {\bibfnamefont {D.}~\bibnamefont {Ashery}}, \bibinfo
  {author} {\bibfnamefont {O.}~\bibnamefont {Aviv}}, \bibinfo {author}
  {\bibfnamefont {J.~A.}\ \bibnamefont {Behr}}, \bibinfo {author}
  {\bibfnamefont {P.~G.}\ \bibnamefont {Bricault}}, \bibinfo {author}
  {\bibfnamefont {M.}~\bibnamefont {Dombsky}}, \bibinfo {author} {\bibfnamefont
  {J.~D.}\ \bibnamefont {Holt}}, \bibinfo {author} {\bibfnamefont {K.~P.}\
  \bibnamefont {Jackson}}, \bibinfo {author} {\bibfnamefont {B.}~\bibnamefont
  {Lee}}, \bibinfo {author} {\bibfnamefont {M.~R.}\ \bibnamefont {Pearson}},
  \bibinfo {author} {\bibfnamefont {A.}~\bibnamefont {Gaudin}}, \bibinfo
  {author} {\bibfnamefont {B.}~\bibnamefont {Dej}}, \bibinfo {author}
  {\bibfnamefont {C.}~\bibnamefont {H\"{o}hr}}, \bibinfo {author}
  {\bibfnamefont {G.}~\bibnamefont {Gwinner}}, \ and\ \bibinfo {author}
  {\bibfnamefont {D.}~\bibnamefont {Melconian}},\ }\href@noop {} {\bibfield
  {journal} {\bibinfo  {journal} {Physical Review C (Nuclear Physics)}\
  }\textbf {\bibinfo {volume} {79}},\ \bibinfo {pages} {015501} (\bibinfo
  {year} {2009})}\BibitemShut {NoStop}%
\bibitem [{\citenamefont {Bolotov}\ \emph {et~al.}(1990)\citenamefont
  {Bolotov}, \citenamefont {Gninenko}, \citenamefont {Djilkibaev},
  \citenamefont {Isakov}, \citenamefont {Klubakov}, \citenamefont {Laptev},
  \citenamefont {Lobashev}, \citenamefont {Marin}, \citenamefont {Poblaguev},
  \citenamefont {Postoev},\ and\ \citenamefont {Toropin}}]{Bolotov:1990}%
  \BibitemOpen
  \bibfield  {author} {\bibinfo {author} {\bibfnamefont {V.}~\bibnamefont
  {Bolotov}}, \bibinfo {author} {\bibfnamefont {S.}~\bibnamefont {Gninenko}},
  \bibinfo {author} {\bibfnamefont {R.}~\bibnamefont {Djilkibaev}}, \bibinfo
  {author} {\bibfnamefont {V.}~\bibnamefont {Isakov}}, \bibinfo {author}
  {\bibfnamefont {Y.}~\bibnamefont {Klubakov}}, \bibinfo {author}
  {\bibfnamefont {V.}~\bibnamefont {Laptev}}, \bibinfo {author} {\bibfnamefont
  {V.}~\bibnamefont {Lobashev}}, \bibinfo {author} {\bibfnamefont
  {V.}~\bibnamefont {Marin}}, \bibinfo {author} {\bibfnamefont
  {A.}~\bibnamefont {Poblaguev}}, \bibinfo {author} {\bibfnamefont
  {V.}~\bibnamefont {Postoev}}, \ and\ \bibinfo {author} {\bibfnamefont
  {A.}~\bibnamefont {Toropin}},\ }\href {\doibase
  http://dx.doi.org/10.1016/0370-2693(90)90857-3} {\bibfield  {journal}
  {\bibinfo  {journal} {Physics Letters B}\ }\textbf {\bibinfo {volume}
  {243}},\ \bibinfo {pages} {308 } (\bibinfo {year} {1990})}\BibitemShut
  {NoStop}%
\bibitem [{\citenamefont {Frlez}\ \emph {et~al.}(2004)\citenamefont {Frlez},
  \citenamefont {Pocanic}, \citenamefont {Baranov}, \citenamefont {Bertl},
  \citenamefont {Bychkov}, \citenamefont {Khomutov}, \citenamefont
  {Korenchenko}, \citenamefont {Korenchenko}, \citenamefont {Kozlowski},
  \citenamefont {Kravchuk}, \citenamefont {Kuchinsky}, \citenamefont {Li},
  \citenamefont {Minehart}, \citenamefont {Mzhavia}, \citenamefont {Ritchie},
  \citenamefont {Ritt}, \citenamefont {Rozhdestvensky}, \citenamefont
  {Sidorkin}, \citenamefont {Smith}, \citenamefont {Supek}, \citenamefont
  {Tsamalaidze}, \citenamefont {VanDevender}, \citenamefont {Velicheva},
  \citenamefont {Wang}, \citenamefont {Wirtz},\ and\ \citenamefont
  {Ziock}}]{Frlez:2004}%
  \BibitemOpen
  \bibfield  {author} {\bibinfo {author} {\bibfnamefont {E.}~\bibnamefont
  {Frlez}}, \bibinfo {author} {\bibfnamefont {D.}~\bibnamefont {Pocanic}},
  \bibinfo {author} {\bibfnamefont {V.~A.}\ \bibnamefont {Baranov}}, \bibinfo
  {author} {\bibfnamefont {W.}~\bibnamefont {Bertl}}, \bibinfo {author}
  {\bibfnamefont {M.}~\bibnamefont {Bychkov}}, \bibinfo {author} {\bibfnamefont
  {N.~V.}\ \bibnamefont {Khomutov}}, \bibinfo {author} {\bibfnamefont {A.~S.}\
  \bibnamefont {Korenchenko}}, \bibinfo {author} {\bibfnamefont {S.~M.}\
  \bibnamefont {Korenchenko}}, \bibinfo {author} {\bibfnamefont
  {T.}~\bibnamefont {Kozlowski}}, \bibinfo {author} {\bibfnamefont {N.~P.}\
  \bibnamefont {Kravchuk}}, \bibinfo {author} {\bibfnamefont {N.~A.}\
  \bibnamefont {Kuchinsky}}, \bibinfo {author} {\bibfnamefont {W.}~\bibnamefont
  {Li}}, \bibinfo {author} {\bibfnamefont {R.~C.}\ \bibnamefont {Minehart}},
  \bibinfo {author} {\bibfnamefont {D.}~\bibnamefont {Mzhavia}}, \bibinfo
  {author} {\bibfnamefont {B.~G.}\ \bibnamefont {Ritchie}}, \bibinfo {author}
  {\bibfnamefont {S.}~\bibnamefont {Ritt}}, \bibinfo {author} {\bibfnamefont
  {A.~M.}\ \bibnamefont {Rozhdestvensky}}, \bibinfo {author} {\bibfnamefont
  {V.~V.}\ \bibnamefont {Sidorkin}}, \bibinfo {author} {\bibfnamefont {L.~C.}\
  \bibnamefont {Smith}}, \bibinfo {author} {\bibfnamefont {I.}~\bibnamefont
  {Supek}}, \bibinfo {author} {\bibfnamefont {Z.}~\bibnamefont {Tsamalaidze}},
  \bibinfo {author} {\bibfnamefont {B.~A.}\ \bibnamefont {VanDevender}},
  \bibinfo {author} {\bibfnamefont {E.~P.}\ \bibnamefont {Velicheva}}, \bibinfo
  {author} {\bibfnamefont {Y.}~\bibnamefont {Wang}}, \bibinfo {author}
  {\bibfnamefont {H.-P.}\ \bibnamefont {Wirtz}}, \ and\ \bibinfo {author}
  {\bibfnamefont {K.~O.~H.}\ \bibnamefont {Ziock}},\ }\href {\doibase
  10.1103/PhysRevLett.93.181804} {\bibfield  {journal} {\bibinfo  {journal}
  {Phys. Rev. Lett.}\ }\textbf {\bibinfo {volume} {93}},\ \bibinfo {pages}
  {181804} (\bibinfo {year} {2004})}\BibitemShut {NoStop}%
\bibitem [{\citenamefont {Poblaguev}(1990)}]{Poblaguev:1990}%
  \BibitemOpen
  \bibfield  {author} {\bibinfo {author} {\bibfnamefont {A.}~\bibnamefont
  {Poblaguev}},\ }\href {\doibase
  http://dx.doi.org/10.1016/0370-2693(90)92108-U} {\bibfield  {journal}
  {\bibinfo  {journal} {Physics Letters B}\ }\textbf {\bibinfo {volume}
  {238}},\ \bibinfo {pages} {108 } (\bibinfo {year} {1990})}\BibitemShut
  {NoStop}%
\bibitem [{\citenamefont {Quin}\ \emph {et~al.}(1993)\citenamefont {Quin},
  \citenamefont {Deutsch}, \citenamefont {Pickering}, \citenamefont {Schewe},\
  and\ \citenamefont {Voytas}}]{Quin1993}%
  \BibitemOpen
  \bibfield  {author} {\bibinfo {author} {\bibfnamefont {P.~A.}\ \bibnamefont
  {Quin}}, \bibinfo {author} {\bibfnamefont {J.}~\bibnamefont {Deutsch}},
  \bibinfo {author} {\bibfnamefont {T.~E.}\ \bibnamefont {Pickering}}, \bibinfo
  {author} {\bibfnamefont {J.~E.}\ \bibnamefont {Schewe}}, \ and\ \bibinfo
  {author} {\bibfnamefont {P.~A.}\ \bibnamefont {Voytas}},\ }\href@noop {}
  {\bibfield  {journal} {\bibinfo  {journal} {Phys. Rev. D}\ }\textbf {\bibinfo
  {volume} {47}},\ \bibinfo {pages} {1247} (\bibinfo {year}
  {1993})}\BibitemShut {NoStop}%
\bibitem [{\citenamefont {Carnoy}\ \emph {et~al.}(1991)\citenamefont {Carnoy},
  \citenamefont {Deutsch}, \citenamefont {Girard},\ and\ \citenamefont
  {Prieels}}]{Carnoy:1991}%
  \BibitemOpen
  \bibfield  {author} {\bibinfo {author} {\bibfnamefont {A.~S.}\ \bibnamefont
  {Carnoy}}, \bibinfo {author} {\bibfnamefont {J.}~\bibnamefont {Deutsch}},
  \bibinfo {author} {\bibfnamefont {T.~A.}\ \bibnamefont {Girard}}, \ and\
  \bibinfo {author} {\bibfnamefont {R.}~\bibnamefont {Prieels}},\ }\href
  {\doibase 10.1103/PhysRevC.43.2825} {\bibfield  {journal} {\bibinfo
  {journal} {Phys. Rev. C}\ }\textbf {\bibinfo {volume} {43}},\ \bibinfo
  {pages} {2825} (\bibinfo {year} {1991})}\BibitemShut {NoStop}%
\bibitem [{\citenamefont {Wichers}\ \emph {et~al.}(1987)\citenamefont
  {Wichers}, \citenamefont {Hageman}, \citenamefont {van Klinken},
  \citenamefont {Wilschut},\ and\ \citenamefont {Atkinson}}]{Wichers:1987}%
  \BibitemOpen
  \bibfield  {author} {\bibinfo {author} {\bibfnamefont {V.~A.}\ \bibnamefont
  {Wichers}}, \bibinfo {author} {\bibfnamefont {T.~R.}\ \bibnamefont
  {Hageman}}, \bibinfo {author} {\bibfnamefont {J.}~\bibnamefont {van
  Klinken}}, \bibinfo {author} {\bibfnamefont {H.~W.}\ \bibnamefont
  {Wilschut}}, \ and\ \bibinfo {author} {\bibfnamefont {D.}~\bibnamefont
  {Atkinson}},\ }\href {\doibase 10.1103/PhysRevLett.58.1821} {\bibfield
  {journal} {\bibinfo  {journal} {Phys. Rev. Lett.}\ }\textbf {\bibinfo
  {volume} {58}},\ \bibinfo {pages} {1821} (\bibinfo {year}
  {1987})}\BibitemShut {NoStop}%
\bibitem [{\citenamefont {Hardy}\ and\ \citenamefont
  {Towner}(2009)}]{Hardy2009}%
  \BibitemOpen
  \bibfield  {author} {\bibinfo {author} {\bibfnamefont {J.~C.}\ \bibnamefont
  {Hardy}}\ and\ \bibinfo {author} {\bibfnamefont {I.~S.}\ \bibnamefont
  {Towner}},\ }\href {\doibase 10.1103/PhysRevC.79.055502} {\bibfield
  {journal} {\bibinfo  {journal} {Physical Review C (Nuclear Physics)}\
  }\textbf {\bibinfo {volume} {79}},\ \bibinfo {eid} {055502} (\bibinfo {year}
  {2009})}\BibitemShut {NoStop}%
\bibitem [{\citenamefont {Plaster}\ \emph {et~al.}(2012)\citenamefont
  {Plaster}, \citenamefont {Rios}, \citenamefont {Back}, \citenamefont
  {Bowles}, \citenamefont {Broussard}, \citenamefont {Carr}, \citenamefont
  {Clayton}, \citenamefont {Currie}, \citenamefont {Filippone}, \citenamefont
  {Garc\'{i}a}, \citenamefont {Geltenbort}, \citenamefont {Hickerson},
  \citenamefont {Hoagland}, \citenamefont {Hogan}, \citenamefont {Hona},
  \citenamefont {Holley}, \citenamefont {Ito}, \citenamefont {Liu},
  \citenamefont {Liu}, \citenamefont {Makela}, \citenamefont {Mammei},
  \citenamefont {Martin}, \citenamefont {Melconian}, \citenamefont
  {Mendenhall}, \citenamefont {Morris}, \citenamefont {Mortensen},
  \citenamefont {Pattie}, \citenamefont {P\'erez~Galv\'an}, \citenamefont
  {Pitt}, \citenamefont {Ramsey}, \citenamefont {Russell}, \citenamefont
  {Saunders}, \citenamefont {Schmid}, \citenamefont {Seestrom}, \citenamefont
  {Sjue}, \citenamefont {Sondheim}, \citenamefont {Tatar}, \citenamefont
  {Tipton}, \citenamefont {Vogelaar}, \citenamefont {VornDick}, \citenamefont
  {Wrede}, \citenamefont {Xu}, \citenamefont {Yan}, \citenamefont {Young},\
  and\ \citenamefont {Yuan}}]{Plaster:2012}%
  \BibitemOpen
  \bibfield  {author} {\bibinfo {author} {\bibfnamefont {B.}~\bibnamefont
  {Plaster}}, \bibinfo {author} {\bibfnamefont {R.}~\bibnamefont {Rios}},
  \bibinfo {author} {\bibfnamefont {H.~O.}\ \bibnamefont {Back}}, \bibinfo
  {author} {\bibfnamefont {T.~J.}\ \bibnamefont {Bowles}}, \bibinfo {author}
  {\bibfnamefont {L.~J.}\ \bibnamefont {Broussard}}, \bibinfo {author}
  {\bibfnamefont {R.}~\bibnamefont {Carr}}, \bibinfo {author} {\bibfnamefont
  {S.}~\bibnamefont {Clayton}}, \bibinfo {author} {\bibfnamefont
  {S.}~\bibnamefont {Currie}}, \bibinfo {author} {\bibfnamefont {B.~W.}\
  \bibnamefont {Filippone}}, \bibinfo {author} {\bibfnamefont {A.}~\bibnamefont
  {Garc\'{i}a}}, \bibinfo {author} {\bibfnamefont {P.}~\bibnamefont
  {Geltenbort}}, \bibinfo {author} {\bibfnamefont {K.~P.}\ \bibnamefont
  {Hickerson}}, \bibinfo {author} {\bibfnamefont {J.}~\bibnamefont {Hoagland}},
  \bibinfo {author} {\bibfnamefont {G.~E.}\ \bibnamefont {Hogan}}, \bibinfo
  {author} {\bibfnamefont {B.}~\bibnamefont {Hona}}, \bibinfo {author}
  {\bibfnamefont {A.~T.}\ \bibnamefont {Holley}}, \bibinfo {author}
  {\bibfnamefont {T.~M.}\ \bibnamefont {Ito}}, \bibinfo {author} {\bibfnamefont
  {C.-Y.}\ \bibnamefont {Liu}}, \bibinfo {author} {\bibfnamefont
  {J.}~\bibnamefont {Liu}}, \bibinfo {author} {\bibfnamefont {M.}~\bibnamefont
  {Makela}}, \bibinfo {author} {\bibfnamefont {R.~R.}\ \bibnamefont {Mammei}},
  \bibinfo {author} {\bibfnamefont {J.~W.}\ \bibnamefont {Martin}}, \bibinfo
  {author} {\bibfnamefont {D.}~\bibnamefont {Melconian}}, \bibinfo {author}
  {\bibfnamefont {M.~P.}\ \bibnamefont {Mendenhall}}, \bibinfo {author}
  {\bibfnamefont {C.~L.}\ \bibnamefont {Morris}}, \bibinfo {author}
  {\bibfnamefont {R.}~\bibnamefont {Mortensen}}, \bibinfo {author}
  {\bibfnamefont {R.~W.}\ \bibnamefont {Pattie}}, \bibinfo {author}
  {\bibfnamefont {A.}~\bibnamefont {P\'erez~Galv\'an}}, \bibinfo {author}
  {\bibfnamefont {M.~L.}\ \bibnamefont {Pitt}}, \bibinfo {author}
  {\bibfnamefont {J.~C.}\ \bibnamefont {Ramsey}}, \bibinfo {author}
  {\bibfnamefont {R.}~\bibnamefont {Russell}}, \bibinfo {author} {\bibfnamefont
  {A.}~\bibnamefont {Saunders}}, \bibinfo {author} {\bibfnamefont
  {R.}~\bibnamefont {Schmid}}, \bibinfo {author} {\bibfnamefont {S.~J.}\
  \bibnamefont {Seestrom}}, \bibinfo {author} {\bibfnamefont {S.}~\bibnamefont
  {Sjue}}, \bibinfo {author} {\bibfnamefont {W.~E.}\ \bibnamefont {Sondheim}},
  \bibinfo {author} {\bibfnamefont {E.}~\bibnamefont {Tatar}}, \bibinfo
  {author} {\bibfnamefont {B.}~\bibnamefont {Tipton}}, \bibinfo {author}
  {\bibfnamefont {R.~B.}\ \bibnamefont {Vogelaar}}, \bibinfo {author}
  {\bibfnamefont {B.}~\bibnamefont {VornDick}}, \bibinfo {author}
  {\bibfnamefont {C.}~\bibnamefont {Wrede}}, \bibinfo {author} {\bibfnamefont
  {Y.~P.}\ \bibnamefont {Xu}}, \bibinfo {author} {\bibfnamefont
  {H.}~\bibnamefont {Yan}}, \bibinfo {author} {\bibfnamefont {A.~R.}\
  \bibnamefont {Young}}, \ and\ \bibinfo {author} {\bibfnamefont
  {J.}~\bibnamefont {Yuan}} (\bibinfo {collaboration} {UCNA Collaboration}),\
  }\href {\doibase 10.1103/PhysRevC.86.055501} {\bibfield  {journal} {\bibinfo
  {journal} {Phys. Rev. C}\ }\textbf {\bibinfo {volume} {86}},\ \bibinfo
  {pages} {055501} (\bibinfo {year} {2012})}\BibitemShut {NoStop}%
\bibitem [{\citenamefont {Abele}\ \emph {et~al.}(2002)\citenamefont {Abele},
  \citenamefont {Hoffmann}, \citenamefont {Bae\ss{}ler}, \citenamefont
  {Dubbers}, \citenamefont {Gl\"uck}, \citenamefont {M\"uller}, \citenamefont
  {Nesvizhevsky}, \citenamefont {Reich},\ and\ \citenamefont
  {Zimmer}}]{Abele2002}%
  \BibitemOpen
  \bibfield  {author} {\bibinfo {author} {\bibfnamefont {H.}~\bibnamefont
  {Abele}}, \bibinfo {author} {\bibfnamefont {M.~A.}\ \bibnamefont {Hoffmann}},
  \bibinfo {author} {\bibfnamefont {S.}~\bibnamefont {Bae\ss{}ler}}, \bibinfo
  {author} {\bibfnamefont {D.}~\bibnamefont {Dubbers}}, \bibinfo {author}
  {\bibfnamefont {F.}~\bibnamefont {Gl\"uck}}, \bibinfo {author} {\bibfnamefont
  {U.}~\bibnamefont {M\"uller}}, \bibinfo {author} {\bibfnamefont
  {V.}~\bibnamefont {Nesvizhevsky}}, \bibinfo {author} {\bibfnamefont
  {J.}~\bibnamefont {Reich}}, \ and\ \bibinfo {author} {\bibfnamefont
  {O.}~\bibnamefont {Zimmer}},\ }\href@noop {} {\bibfield  {journal} {\bibinfo
  {journal} {Phys. Rev. Lett.}\ }\textbf {\bibinfo {volume} {88}},\ \bibinfo
  {pages} {211801} (\bibinfo {year} {2002})}\BibitemShut {NoStop}%
\bibitem [{\citenamefont {Abele}\ \emph {et~al.}(1997)\citenamefont {Abele},
  \citenamefont {Baeßler}, \citenamefont {Dubbers}, \citenamefont {Last},
  \citenamefont {Mayerhofer}, \citenamefont {Metz}, \citenamefont {Müller},
  \citenamefont {Nesvizhevsky}, \citenamefont {Raven}, \citenamefont
  {Schärpf},\ and\ \citenamefont {Zimmer}}]{Abele1997}%
  \BibitemOpen
  \bibfield  {author} {\bibinfo {author} {\bibfnamefont {H.}~\bibnamefont
  {Abele}}, \bibinfo {author} {\bibfnamefont {S.}~\bibnamefont {Baeßler}},
  \bibinfo {author} {\bibfnamefont {D.}~\bibnamefont {Dubbers}}, \bibinfo
  {author} {\bibfnamefont {J.}~\bibnamefont {Last}}, \bibinfo {author}
  {\bibfnamefont {U.}~\bibnamefont {Mayerhofer}}, \bibinfo {author}
  {\bibfnamefont {C.}~\bibnamefont {Metz}}, \bibinfo {author} {\bibfnamefont
  {T.}~\bibnamefont {Müller}}, \bibinfo {author} {\bibfnamefont
  {V.}~\bibnamefont {Nesvizhevsky}}, \bibinfo {author} {\bibfnamefont
  {C.}~\bibnamefont {Raven}}, \bibinfo {author} {\bibfnamefont
  {O.}~\bibnamefont {Schärpf}}, \ and\ \bibinfo {author} {\bibfnamefont
  {O.}~\bibnamefont {Zimmer}},\ }\href
  {http://www.sciencedirect.com/science/article/pii/S0370269397007399}
  {\bibfield  {journal} {\bibinfo  {journal} {Physics Letters B}\ }\textbf
  {\bibinfo {volume} {407}},\ \bibinfo {pages} {212 } (\bibinfo {year}
  {1997})}\BibitemShut {NoStop}%
\bibitem [{\citenamefont {Liaud}\ \emph {et~al.}(1997)\citenamefont {Liaud},
  \citenamefont {Schreckenbach}, \citenamefont {Kossakowski}, \citenamefont
  {Nastoll}, \citenamefont {Bussière}, \citenamefont {Guillaud},\ and\
  \citenamefont {Beck}}]{Liaud1997}%
  \BibitemOpen
  \bibfield  {author} {\bibinfo {author} {\bibfnamefont {P.}~\bibnamefont
  {Liaud}}, \bibinfo {author} {\bibfnamefont {K.}~\bibnamefont
  {Schreckenbach}}, \bibinfo {author} {\bibfnamefont {R.}~\bibnamefont
  {Kossakowski}}, \bibinfo {author} {\bibfnamefont {H.}~\bibnamefont
  {Nastoll}}, \bibinfo {author} {\bibfnamefont {A.}~\bibnamefont {Bussière}},
  \bibinfo {author} {\bibfnamefont {J.}~\bibnamefont {Guillaud}}, \ and\
  \bibinfo {author} {\bibfnamefont {L.}~\bibnamefont {Beck}},\ }\href@noop {}
  {\bibfield  {journal} {\bibinfo  {journal} {Nuclear Physics A}\ }\textbf
  {\bibinfo {volume} {612}},\ \bibinfo {pages} {53 } (\bibinfo {year}
  {1997})}\BibitemShut {NoStop}%
\bibitem [{\citenamefont {Yerozolimsky}\ \emph {et~al.}(1997)\citenamefont
  {Yerozolimsky}, \citenamefont {Kuznetsov}, \citenamefont {Mostovoy},\ and\
  \citenamefont {Stepanenko}}]{Yerozolimsky1997}%
  \BibitemOpen
  \bibfield  {author} {\bibinfo {author} {\bibfnamefont {B.}~\bibnamefont
  {Yerozolimsky}}, \bibinfo {author} {\bibfnamefont {I.}~\bibnamefont
  {Kuznetsov}}, \bibinfo {author} {\bibfnamefont {Y.}~\bibnamefont {Mostovoy}},
  \ and\ \bibinfo {author} {\bibfnamefont {I.}~\bibnamefont {Stepanenko}},\
  }\href@noop {} {\bibfield  {journal} {\bibinfo  {journal} {Physics Letters
  B}\ }\textbf {\bibinfo {volume} {412}},\ \bibinfo {pages} {240 } (\bibinfo
  {year} {1997})}\BibitemShut {NoStop}%
\bibitem [{\citenamefont {Erozolimsky}\ \emph {et~al.}(1991)\citenamefont
  {Erozolimsky}, \citenamefont {Kuznetsov}, \citenamefont {Kujda},
  \citenamefont {Mostovoi},\ and\ \citenamefont
  {Stepanenko}}]{Erozolimsky:1990}%
  \BibitemOpen
  \bibfield  {author} {\bibinfo {author} {\bibfnamefont {B.}~\bibnamefont
  {Erozolimsky}}, \bibinfo {author} {\bibfnamefont {I.}~\bibnamefont
  {Kuznetsov}}, \bibinfo {author} {\bibfnamefont {I.}~\bibnamefont {Kujda}},
  \bibinfo {author} {\bibfnamefont {Y.}~\bibnamefont {Mostovoi}}, \ and\
  \bibinfo {author} {\bibfnamefont {I.}~\bibnamefont {Stepanenko}},\
  }\href@noop {} {\bibfield  {journal} {\bibinfo  {journal} {Phys.Lett.}\
  }\textbf {\bibinfo {volume} {B263}},\ \bibinfo {pages} {33} (\bibinfo {year}
  {1991})}\BibitemShut {NoStop}%
\bibitem [{\citenamefont {Bopp}\ \emph {et~al.}(1986)\citenamefont {Bopp},
  \citenamefont {Dubbers}, \citenamefont {Hornig}, \citenamefont {Klemt},
  \citenamefont {Last}, \citenamefont {Sch\"utze}, \citenamefont {Freedman},\
  and\ \citenamefont {Sch\"arpf}}]{Bopp1986}%
  \BibitemOpen
  \bibfield  {author} {\bibinfo {author} {\bibfnamefont {P.}~\bibnamefont
  {Bopp}}, \bibinfo {author} {\bibfnamefont {D.}~\bibnamefont {Dubbers}},
  \bibinfo {author} {\bibfnamefont {L.}~\bibnamefont {Hornig}}, \bibinfo
  {author} {\bibfnamefont {E.}~\bibnamefont {Klemt}}, \bibinfo {author}
  {\bibfnamefont {J.}~\bibnamefont {Last}}, \bibinfo {author} {\bibfnamefont
  {H.}~\bibnamefont {Sch\"utze}}, \bibinfo {author} {\bibfnamefont {S.~J.}\
  \bibnamefont {Freedman}}, \ and\ \bibinfo {author} {\bibfnamefont
  {O.}~\bibnamefont {Sch\"arpf}},\ }\href {\doibase 10.1103/PhysRevLett.56.919}
  {\bibfield  {journal} {\bibinfo  {journal} {Phys. Rev. Lett.}\ }\textbf
  {\bibinfo {volume} {56}},\ \bibinfo {pages} {919} (\bibinfo {year}
  {1986})}\BibitemShut {NoStop}%
\bibitem [{\citenamefont {Arzumanov}\ \emph {et~al.}(2012)\citenamefont
  {Arzumanov}, \citenamefont {Bondarenko}, \citenamefont {Morozov},
  \citenamefont {Panin},\ and\ \citenamefont {Chernyavsky}}]{Arzumanov:2012}%
  \BibitemOpen
  \bibfield  {author} {\bibinfo {author} {\bibfnamefont {S.}~\bibnamefont
  {Arzumanov}}, \bibinfo {author} {\bibfnamefont {L.}~\bibnamefont
  {Bondarenko}}, \bibinfo {author} {\bibfnamefont {V.}~\bibnamefont {Morozov}},
  \bibinfo {author} {\bibfnamefont {Y.}~\bibnamefont {Panin}}, \ and\ \bibinfo
  {author} {\bibfnamefont {S.}~\bibnamefont {Chernyavsky}},\ }\href@noop {}
  {\bibfield  {journal} {\bibinfo  {journal} {JETP Lett.}\ }\textbf {\bibinfo
  {volume} {95}},\ \bibinfo {pages} {224} (\bibinfo {year} {2012})}\BibitemShut
  {NoStop}%
\bibitem [{\citenamefont {Arzumanov}\ \emph {et~al.}(2000)\citenamefont
  {Arzumanov}, \citenamefont {Bondarenko}, \citenamefont {Chernyavsky},
  \citenamefont {Drexel}, \citenamefont {Fomin}, \citenamefont {Geltenbort},
  \citenamefont {Morozov}, \citenamefont {Panin}, \citenamefont {Pendlebury},\
  and\ \citenamefont {Schreckenbach}}]{Arzumanov2000}%
  \BibitemOpen
  \bibfield  {author} {\bibinfo {author} {\bibfnamefont {S.}~\bibnamefont
  {Arzumanov}}, \bibinfo {author} {\bibfnamefont {L.}~\bibnamefont
  {Bondarenko}}, \bibinfo {author} {\bibfnamefont {S.}~\bibnamefont
  {Chernyavsky}}, \bibinfo {author} {\bibfnamefont {W.}~\bibnamefont {Drexel}},
  \bibinfo {author} {\bibfnamefont {A.}~\bibnamefont {Fomin}}, \bibinfo
  {author} {\bibfnamefont {P.}~\bibnamefont {Geltenbort}}, \bibinfo {author}
  {\bibfnamefont {V.}~\bibnamefont {Morozov}}, \bibinfo {author} {\bibfnamefont
  {Y.}~\bibnamefont {Panin}}, \bibinfo {author} {\bibfnamefont
  {J.}~\bibnamefont {Pendlebury}}, \ and\ \bibinfo {author} {\bibfnamefont
  {K.}~\bibnamefont {Schreckenbach}},\ }\href@noop {} {\bibfield  {journal}
  {\bibinfo  {journal} {Physics Letters B}\ }\textbf {\bibinfo {volume}
  {483}},\ \bibinfo {pages} {15 } (\bibinfo {year} {2000})}\BibitemShut
  {NoStop}%
\bibitem [{\citenamefont {Pichlmaier}\ \emph {et~al.}(2010)\citenamefont
  {Pichlmaier}, \citenamefont {Varlamov}, \citenamefont {Schreckenbach},\ and\
  \citenamefont {Geltenbort}}]{Pichlmaier:2010}%
  \BibitemOpen
  \bibfield  {author} {\bibinfo {author} {\bibfnamefont {A.}~\bibnamefont
  {Pichlmaier}}, \bibinfo {author} {\bibfnamefont {V.}~\bibnamefont
  {Varlamov}}, \bibinfo {author} {\bibfnamefont {K.}~\bibnamefont
  {Schreckenbach}}, \ and\ \bibinfo {author} {\bibfnamefont {P.}~\bibnamefont
  {Geltenbort}},\ }\href@noop {} {\bibfield  {journal} {\bibinfo  {journal}
  {Phys.Lett.}\ }\textbf {\bibinfo {volume} {B 693}},\ \bibinfo {pages} {221}
  (\bibinfo {year} {2010})}\BibitemShut {NoStop}%
\bibitem [{\citenamefont {Nico}\ \emph {et~al.}(2005)\citenamefont {Nico},
  \citenamefont {Dewey}, \citenamefont {Gilliam}, \citenamefont {Wietfeldt},
  \citenamefont {Fei}, \citenamefont {Snow}, \citenamefont {Greene},
  \citenamefont {Pauwels}, \citenamefont {Eykens}, \citenamefont {Lamberty},
  \citenamefont {Van~Gestel},\ and\ \citenamefont {Scott}}]{NICO2005}%
  \BibitemOpen
  \bibfield  {author} {\bibinfo {author} {\bibfnamefont {J.~S.}\ \bibnamefont
  {Nico}}, \bibinfo {author} {\bibfnamefont {M.~S.}\ \bibnamefont {Dewey}},
  \bibinfo {author} {\bibfnamefont {D.~M.}\ \bibnamefont {Gilliam}}, \bibinfo
  {author} {\bibfnamefont {F.~E.}\ \bibnamefont {Wietfeldt}}, \bibinfo {author}
  {\bibfnamefont {X.}~\bibnamefont {Fei}}, \bibinfo {author} {\bibfnamefont
  {W.~M.}\ \bibnamefont {Snow}}, \bibinfo {author} {\bibfnamefont {G.~L.}\
  \bibnamefont {Greene}}, \bibinfo {author} {\bibfnamefont {J.}~\bibnamefont
  {Pauwels}}, \bibinfo {author} {\bibfnamefont {R.}~\bibnamefont {Eykens}},
  \bibinfo {author} {\bibfnamefont {A.}~\bibnamefont {Lamberty}}, \bibinfo
  {author} {\bibfnamefont {J.}~\bibnamefont {Van~Gestel}}, \ and\ \bibinfo
  {author} {\bibfnamefont {R.~D.}\ \bibnamefont {Scott}},\ }\href@noop {}
  {\bibfield  {journal} {\bibinfo  {journal} {Phys. Rev. C}\ }\textbf {\bibinfo
  {volume} {71}},\ \bibinfo {pages} {055502} (\bibinfo {year}
  {2005})}\BibitemShut {NoStop}%
\bibitem [{\citenamefont {Serebrov}\ \emph {et~al.}(2005)\citenamefont
  {Serebrov}, \citenamefont {Varlamov}, \citenamefont {Kharitonov},
  \citenamefont {Fomin}, \citenamefont {Pokotilovski}, \citenamefont
  {Geltenbort}, \citenamefont {Butterworth}, \citenamefont {Krasnoschekova},
  \citenamefont {Lasakov}, \citenamefont {Tal'daev}, \citenamefont
  {Vassiljev},\ and\ \citenamefont {Zherebtsov}}]{Serebrov2005}%
  \BibitemOpen
  \bibfield  {author} {\bibinfo {author} {\bibfnamefont {A.}~\bibnamefont
  {Serebrov}}, \bibinfo {author} {\bibfnamefont {V.}~\bibnamefont {Varlamov}},
  \bibinfo {author} {\bibfnamefont {A.}~\bibnamefont {Kharitonov}}, \bibinfo
  {author} {\bibfnamefont {A.}~\bibnamefont {Fomin}}, \bibinfo {author}
  {\bibfnamefont {Y.}~\bibnamefont {Pokotilovski}}, \bibinfo {author}
  {\bibfnamefont {P.}~\bibnamefont {Geltenbort}}, \bibinfo {author}
  {\bibfnamefont {J.}~\bibnamefont {Butterworth}}, \bibinfo {author}
  {\bibfnamefont {I.}~\bibnamefont {Krasnoschekova}}, \bibinfo {author}
  {\bibfnamefont {M.}~\bibnamefont {Lasakov}}, \bibinfo {author} {\bibfnamefont
  {R.}~\bibnamefont {Tal'daev}}, \bibinfo {author} {\bibfnamefont
  {A.}~\bibnamefont {Vassiljev}}, \ and\ \bibinfo {author} {\bibfnamefont
  {O.}~\bibnamefont {Zherebtsov}},\ }\href@noop {} {\bibfield  {journal}
  {\bibinfo  {journal} {Physics Letters B}\ }\textbf {\bibinfo {volume}
  {605}},\ \bibinfo {pages} {72 } (\bibinfo {year} {2005})}\BibitemShut
  {NoStop}%
\bibitem [{\citenamefont {Byrne}\ and\ \citenamefont
  {Dawber}(1996)}]{Byrne:1996}%
  \BibitemOpen
  \bibfield  {author} {\bibinfo {author} {\bibfnamefont {J.}~\bibnamefont
  {Byrne}}\ and\ \bibinfo {author} {\bibfnamefont {P.}~\bibnamefont {Dawber}},\
  }\href@noop {} {\bibfield  {journal} {\bibinfo  {journal} {Europhys.Lett.}\
  }\textbf {\bibinfo {volume} {33}},\ \bibinfo {pages} {187} (\bibinfo {year}
  {1996})}\BibitemShut {NoStop}%
\bibitem [{\citenamefont {Mampe}\ \emph {et~al.}(1993)\citenamefont {Mampe},
  \citenamefont {Bondarenko}, \citenamefont {Morozov}, \citenamefont {Panin},\
  and\ \citenamefont {Fomin}}]{Mampe:1993}%
  \BibitemOpen
  \bibfield  {author} {\bibinfo {author} {\bibfnamefont {W.}~\bibnamefont
  {Mampe}}, \bibinfo {author} {\bibfnamefont {L.}~\bibnamefont {Bondarenko}},
  \bibinfo {author} {\bibfnamefont {V.}~\bibnamefont {Morozov}}, \bibinfo
  {author} {\bibfnamefont {Y.}~\bibnamefont {Panin}}, \ and\ \bibinfo {author}
  {\bibfnamefont {A.}~\bibnamefont {Fomin}},\ }\href@noop {} {\bibfield
  {journal} {\bibinfo  {journal} {JETP Lett.}\ }\textbf {\bibinfo {volume}
  {57}},\ \bibinfo {pages} {82} (\bibinfo {year} {1993})}\BibitemShut {NoStop}%
\bibitem [{\citenamefont {Mampe}\ \emph {et~al.}(1989)\citenamefont {Mampe},
  \citenamefont {Ageron}, \citenamefont {Bates}, \citenamefont {Pendlebury},\
  and\ \citenamefont {Steyerl}}]{Mampe:1989}%
  \BibitemOpen
  \bibfield  {author} {\bibinfo {author} {\bibfnamefont {W.}~\bibnamefont
  {Mampe}}, \bibinfo {author} {\bibfnamefont {P.}~\bibnamefont {Ageron}},
  \bibinfo {author} {\bibfnamefont {C.}~\bibnamefont {Bates}}, \bibinfo
  {author} {\bibfnamefont {J.~M.}\ \bibnamefont {Pendlebury}}, \ and\ \bibinfo
  {author} {\bibfnamefont {A.}~\bibnamefont {Steyerl}},\ }\href@noop {}
  {\bibfield  {journal} {\bibinfo  {journal} {Phys.Rev.Lett.}\ }\textbf
  {\bibinfo {volume} {63}},\ \bibinfo {pages} {593} (\bibinfo {year}
  {1989})}\BibitemShut {NoStop}%
\bibitem [{\citenamefont {Knecht}\ \emph {et~al.}(2011)\citenamefont {Knecht},
  \citenamefont {Zumwalt}, \citenamefont {Delbridge}, \citenamefont {Garcia},
  \citenamefont {Harper}, \citenamefont {Hong}, \citenamefont {Mueller},
  \citenamefont {Palmer}, \citenamefont {Robertson}, \citenamefont {Swanson},
  \citenamefont {Utsuno}, \citenamefont {Will}, \citenamefont {Williams},\ and\
  \citenamefont {Wrede}}]{Knecht2011}%
  \BibitemOpen
  \bibfield  {author} {\bibinfo {author} {\bibfnamefont {A.}~\bibnamefont
  {Knecht}}, \bibinfo {author} {\bibfnamefont {D.~W.}\ \bibnamefont {Zumwalt}},
  \bibinfo {author} {\bibfnamefont {B.~G.}\ \bibnamefont {Delbridge}}, \bibinfo
  {author} {\bibfnamefont {A.}~\bibnamefont {Garcia}}, \bibinfo {author}
  {\bibfnamefont {G.~C.}\ \bibnamefont {Harper}}, \bibinfo {author}
  {\bibfnamefont {R.}~\bibnamefont {Hong}}, \bibinfo {author} {\bibfnamefont
  {P.}~\bibnamefont {Mueller}}, \bibinfo {author} {\bibfnamefont {A.~S.~C.}\
  \bibnamefont {Palmer}}, \bibinfo {author} {\bibfnamefont {R.~G.~H.}\
  \bibnamefont {Robertson}}, \bibinfo {author} {\bibfnamefont {H.~E.}\
  \bibnamefont {Swanson}}, \bibinfo {author} {\bibfnamefont {S.}~\bibnamefont
  {Utsuno}}, \bibinfo {author} {\bibfnamefont {D.~I.}\ \bibnamefont {Will}},
  \bibinfo {author} {\bibfnamefont {W.}~\bibnamefont {Williams}}, \ and\
  \bibinfo {author} {\bibfnamefont {C.}~\bibnamefont {Wrede}},\ }\href@noop {}
  {\bibfield  {journal} {\bibinfo  {journal} {Nuclear Instruments and Methods
  A}\ }\textbf {\bibinfo {volume} {660}},\ \bibinfo {pages} {43} (\bibinfo
  {year} {2011})}\BibitemShut {NoStop}%
\bibitem [{\citenamefont {Pocanic}\ \emph {et~al.}(2009)\citenamefont
  {Pocanic}, \citenamefont {Alarcon}, \citenamefont {Alonzi}, \citenamefont
  {Baeßler}, \citenamefont {Balascuta}, \citenamefont {Bowman}, \citenamefont
  {Bychkov}, \citenamefont {Byrne}, \citenamefont {Calarco}, \citenamefont
  {Cianciolo}, \citenamefont {Crawford}, \citenamefont {Frlež}, \citenamefont
  {Gericke}, \citenamefont {Greene}, \citenamefont {Grzywacz}, \citenamefont
  {Gudkov}, \citenamefont {Hersman}, \citenamefont {Klein}, \citenamefont
  {Martin}, \citenamefont {Page}, \citenamefont {Palladino}, \citenamefont
  {PenttilA}, \citenamefont {Rykaczewski}, \citenamefont {Wilburn},
  \citenamefont {Young},\ and\ \citenamefont {Young}}]{Pocanic2009}%
  \BibitemOpen
  \bibfield  {author} {\bibinfo {author} {\bibfnamefont {D.}~\bibnamefont
  {Pocanic}}, \bibinfo {author} {\bibfnamefont {R.}~\bibnamefont {Alarcon}},
  \bibinfo {author} {\bibfnamefont {L.}~\bibnamefont {Alonzi}}, \bibinfo
  {author} {\bibfnamefont {S.}~\bibnamefont {Baeßler}}, \bibinfo {author}
  {\bibfnamefont {S.}~\bibnamefont {Balascuta}}, \bibinfo {author}
  {\bibfnamefont {J.}~\bibnamefont {Bowman}}, \bibinfo {author} {\bibfnamefont
  {M.}~\bibnamefont {Bychkov}}, \bibinfo {author} {\bibfnamefont
  {J.}~\bibnamefont {Byrne}}, \bibinfo {author} {\bibfnamefont
  {J.}~\bibnamefont {Calarco}}, \bibinfo {author} {\bibfnamefont
  {V.}~\bibnamefont {Cianciolo}}, \bibinfo {author} {\bibfnamefont
  {C.}~\bibnamefont {Crawford}}, \bibinfo {author} {\bibfnamefont
  {E.}~\bibnamefont {Frlež}}, \bibinfo {author} {\bibfnamefont
  {M.}~\bibnamefont {Gericke}}, \bibinfo {author} {\bibfnamefont
  {G.}~\bibnamefont {Greene}}, \bibinfo {author} {\bibfnamefont
  {R.}~\bibnamefont {Grzywacz}}, \bibinfo {author} {\bibfnamefont
  {V.}~\bibnamefont {Gudkov}}, \bibinfo {author} {\bibfnamefont
  {F.}~\bibnamefont {Hersman}}, \bibinfo {author} {\bibfnamefont
  {A.}~\bibnamefont {Klein}}, \bibinfo {author} {\bibfnamefont
  {J.}~\bibnamefont {Martin}}, \bibinfo {author} {\bibfnamefont
  {S.}~\bibnamefont {Page}}, \bibinfo {author} {\bibfnamefont {A.}~\bibnamefont
  {Palladino}}, \bibinfo {author} {\bibfnamefont {S.}~\bibnamefont {PenttilA}},
  \bibinfo {author} {\bibfnamefont {K.}~\bibnamefont {Rykaczewski}}, \bibinfo
  {author} {\bibfnamefont {W.}~\bibnamefont {Wilburn}}, \bibinfo {author}
  {\bibfnamefont {A.}~\bibnamefont {Young}}, \ and\ \bibinfo {author}
  {\bibfnamefont {G.}~\bibnamefont {Young}},\ }\href@noop {} {\bibfield
  {journal} {\bibinfo  {journal} {Nuclear Instruments and Methods in Physics
  Research Section A: Accelerators, Spectrometers, Detectors and Associated
  Equipment}\ }\textbf {\bibinfo {volume} {611}},\ \bibinfo {pages} {211 }
  (\bibinfo {year} {2009})}\BibitemShut {NoStop}%
\bibitem [{\citenamefont {Boothroyd}\ \emph {et~al.}(1984)\citenamefont
  {Boothroyd}, \citenamefont {Markey},\ and\ \citenamefont
  {Vogel}}]{Boothroyd:1984}%
  \BibitemOpen
  \bibfield  {author} {\bibinfo {author} {\bibfnamefont {A.~I.}\ \bibnamefont
  {Boothroyd}}, \bibinfo {author} {\bibfnamefont {J.}~\bibnamefont {Markey}}, \
  and\ \bibinfo {author} {\bibfnamefont {P.}~\bibnamefont {Vogel}},\ }\href
  {\doibase 10.1103/PhysRevC.29.603} {\bibfield  {journal} {\bibinfo  {journal}
  {Phys. Rev. C}\ }\textbf {\bibinfo {volume} {29}},\ \bibinfo {pages} {603}
  (\bibinfo {year} {1984})}\BibitemShut {NoStop}%
\bibitem [{\citenamefont {Gonz\'{a}lez-Alonso}\ and\ \citenamefont
  {O}(2013)}]{Gonzalez2013}%
  \BibitemOpen
  \bibfield  {author} {\bibinfo {author} {\bibfnamefont {M.}~\bibnamefont
  {Gonz\'{a}lez-Alonso}}\ and\ \bibinfo {author} {\bibfnamefont {N.-C.}\
  \bibnamefont {O}},\ }\href {http://arxiv.org/abs/1304.1759} {\bibfield
  {journal} {\bibinfo  {journal} {arXiv:1304.1759 [hep-ph]}\ } (\bibinfo {year}
  {2013})}\BibitemShut {NoStop}%
\end{thebibliography}%

\end{document}